\documentclass[twocolumn,showpacs,preprintnumbers,amsmath,amssymb,floatfix,nofootinbib]{revtex4-1}
\usepackage{graphicx}
\usepackage{bm}
\usepackage{graphicx}
\usepackage{epstopdf}
\usepackage{xcolor}
\usepackage{mathrsfs}

\newcommand{\argmin}{\operatornamewithlimits{arg\,min}}

\newcommand{\argmax}{\operatornamewithlimits{arg\,max}}

\begin{document}

\title{Covariant Active-Hydrodynamics of Shape-Changing Epithelial Monolayers}

\author{Richard G.~Morris and Madan Rao}

\affiliation{Simons Centre for the Study of Living Machines, National Centre 
	for Biological Sciences, \\Tata Institute for Fundamental Research, 
	Bangalore, 560065, India.
}

\begin{abstract}
	During the early-stages of embryo development, morphogenesis--- the 
	emergence of shape and form in living organisms--- is almost exclusively 
	associated with monolayers of tightly bound epithelial cells.  To 
	understand how such tissues change their shape, we construct a fully 
	covariant active-hydrodynamic theory.  At the cellular scale, stresses 
	arise from apical contractility, mechanical response and the constraint of 
	constant cell volume.  Tissue-scale deformations emerge due to the balance 
	between such cell-autonomous stresses and the
	displacement and shear of a low Reynolds number embedding fluid.  
	Tissues with arbitrary curvature or shape can be described, providing a 
	general framework for epithelial monolayer morphology.  Analysis of the 
	stability of flat monolayers reveals two generic shape instabilities: 
	passive constrained-buckling, and actively-driven tissue deformation.  
	The active instability can be further categorised into two types, 
	corresponding to cell shape changes that are either ``squamous to 
	columnar'' or ``regular-prism to truncated-pyramid''.  The deformations 
	resulting from the latter qualitatively reproduce {\it in vivo} 
	observations of the onset of both mesoderm and posterior midgut 
	invaginations, which take place during gastrulation in the fruit fly 
	{\it Drosophila melanogaster}.
\end{abstract}


\maketitle

\begin{figure}[!t]
\centering
\includegraphics[width=0.45\textwidth]{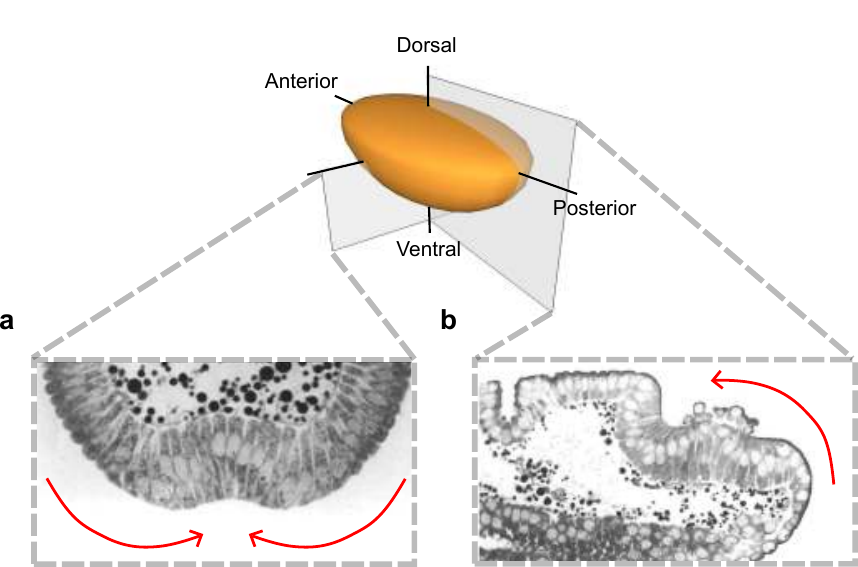}
%/home/rgm/Dropbox/Membranes/Epithelia/Elastomer/Figures/epith_3.pdf
\caption
{
	{\bf Shape-changing tissues.} Gastrulation in the fruit fly {\it Drosophila 
	melanogaster} is characterised by two canonical invaginations, where 
	epithelial monolayers autonomously change their shape with unerring 
	accuracy and robustness.  The mesoderm invagination spans the 
	anterior-posterior axis on the ventral side ({\bf a}), whilst the midgut 
	invagination begins at the posterior pole and progresses towards the 
	anterior along the dorsal side ({\bf b}).  Bright-field images are 
	reproduced with permission from \cite{Sweeton1991}.
}
\label{fig:opening}
\end{figure}

Morphogenesis--- the autonomous formation of shape and form--- is a profound 
process that has fascinated scientists since long before d'Arcy Thompson's 
treatise on the subject one hundred years ago \cite{Thompson1917}.  In the 
context of early-embryo development, the archetypal tissues responsible for 
such deformations are aggregates of epithelial cells, typically arranged in 
thin sheets \cite{Guillot2013a}.  One of the most striking examples, and a 
system subject to intense experimental research, is the process of gastrulation 
in the fruit fly {\it Drosophila melanogaster} \cite{Sweeton1991,Gilmour2017}.  
Here, the epithelium--- a monolayer of cells, tightly connected to each other 
via proteins such as E-cadherin--- undergoes embryo-scale deformations which 
form the basis of the fly's anatomy \cite{BA+89}.  We focus on the mesoderm and 
posterior midgut invaginations (see Fig.~\ref{fig:opening}) where shape changes 
arise as a result of constricting the apical surface of the individual 
epithelial cells; itself attributed to the action of myosin-II motors on an 
especially dense cortical layer of actin that underpins the apical surface 
\cite{Sherrard2010,Lecuit2011,He2014}.  On the timescale of such deformations, 
there is neither proliferation nor cell death, and cells retain the same 
neighbours.  The latter prohibits both the flow of cells relative to each other 
and actively-mediated topological changes, such as the T1-transition.

Our approach is to use a coarse-grained active-hydrodynamic description (in the 
generalised sense of \cite{Marchetti2013,Ramaswamy2010a}), where cells are 
``microscopic'' quantities.  Here, whether relating to gastrulation or some 
other aspect of developmental morphogenesis, changes in tissue geometry are a 
defining (and inescapable) feature which must be characterised.  As a result, 
our treatment is necessarily covariant; cast in the language of differential 
geometry.  A key assumption is to exploit the fact that the lateral scale of 
epithelial monolayers is an order of magnitude larger than the 
thickness\footnote{The semi-major and semi-minor axis of the embryo have 
typical lengths $\sim 200$ $\mu m$ and $\sim150$ $\mu m$, respectively.  The 
typical lateral thickness of the epithelial cells are $\sim10$ $\mu m$.} and 
invoke a ``thin film'' approximation, where the epithelium is represented by a 
single time-dependent manifold that separates two identical semi-infinite 
low-Reynold-number fluids.  The theoretical context is therefore the body of 
work that spans deformations of passive fluid membranes \cite{WCTCL95,MAAD09} 
and, more recently, {\it active} membranes \cite{Maitra2014,Salbreux2017}.

%(similar to methods used in the study of fluid bilayer membranes 
%\cite{PETROV1976a,US96}).

In the embryonic setting, forces generated by the epithelium are balanced by 
the displacement and shear of the highly viscous yolk, or embedding fluid, 
which is the dominant method of dissipation [see \cite{Wirtz2009a,Wessel2015a} 
and Supporting Information (SI)].  Such forces are typically functions of local 
geometry ({\it e.g.}, stretching or bending) and have both passive and active 
contributions.  The former is written in terms of an effective free-energy 
density that captures the mechanical response of cells ({\it e.g.}, apical, 
basal and lateral faces, enclosing an incompressible volume 
\cite{Hannezo2014}).  The latter is assumed to be generated by the ubiquitous 
machinery of the contractile acto-myosin cortex, therefore requiring an 
auxiliary equation for the dynamics of an excitable scalar field for the 
density of myosin-II motors bound to each cell cortex.

In the following, we expand on the above and present a quantitative framework 
to address the leading-order physics of tissue shape changes.  Our closed-form 
theory is, in principle, very general, and able to describe arbitrary shape 
changes.  In practice, however, the treatment of complex geometries and 
non-linearities typically requires a numerical implementation.  We therefore 
validate our framework by analysing the stability of flat steady-state 
(contractile) monolayers.  Notably, even such an ostensibly straightforward 
setting gives rise to a rich set of behaviours, including a novel passive {\it 
constrained}-buckling, and an active {\it invaginating} instability that is 
reminiscent of the onset of both mesoderm and midgut invaginations seen during 
gastrulation of the model organism {\it Drosophila Melanogaster}.

\section{Morphology}
\begin{figure*}[!t]
\centering
\includegraphics[width=0.98\textwidth]{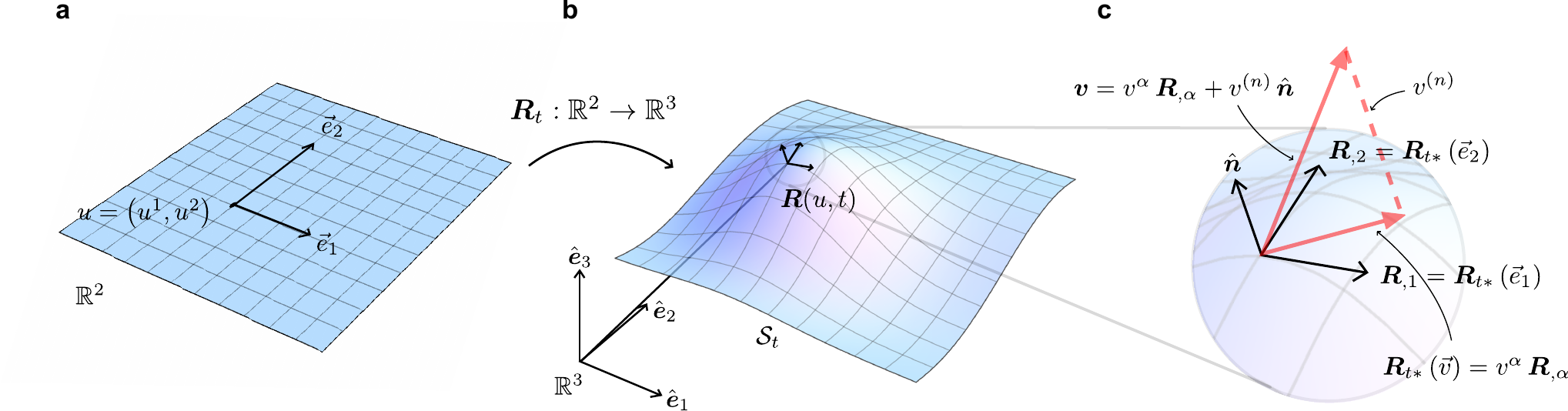}
%/home/rgm/Dropbox/Membranes/Epithelia/Elastomer/Figures/setup_3.pdf
\caption
{
	{\bf Deforming manifolds.} The surface of connected apical faces is 
	parameterised by an Lagrangian coordinate $u\in\mathbb{R}^2$, whose domain 
	is spanned by $\vec{e}_\alpha$ ({\bf a}).  Under the time-dependent 
	embedding $\boldsymbol{R}_t:\mathbb{R}^2\to\mathbb{R}^3$, the positions of 
	points $\boldsymbol{R}(u,t)\in\mathbb{R}^3$ form the manifold 
	$\mathcal{S}_t$ ({\bf b}).  At a given point, the tangent space 
	$T_{\boldsymbol{R}(u,t)}\mathcal{S}_t$ is spanned by 
	$\boldsymbol{R}_{t\,\ast}\left(\vec{e}_\alpha\right)=\boldsymbol{R}_{,\alpha}$.  
	Under deformation, the local velocity $\boldsymbol{v} = \partial 
	\boldsymbol{R}(u,t) / \partial t$ can be decomposed into tangential and 
	normal components: $\boldsymbol{v}=v^\alpha\,\boldsymbol{R}_{,\alpha} + 
	v^{(n)}\hat{\boldsymbol{n}}$ ({\bf c}).
}
\label{fig:setup}
\end{figure*}

Focussing first on the surface of connected apical faces, we use an 
``internal'', Lagrangian, coordinate $u\in\mathbb{R}^2$ to label fixed points 
({\it e.g.}, a given junction between three cells).  The {\it positions} of 
these points $\boldsymbol{R}(u,t)\in\mathbb{R}^3$ then form a manifold 
$\mathcal{S}_t$, which is just the image of $u$ under a time-dependent 
embedding, $\boldsymbol{R}_t:\mathbb{R}^2\to \mathbb{R}^3$ (see 
Fig.~\ref{fig:setup}).  It is the structure of $\mathcal{S}_t$, induced on the 
domain $\mathbb{R}^2$ of $u$, which is of interest.  In particular, the metric, 
$g_{\alpha\beta}(u,t)=\boldsymbol{R}_{,\alpha} \cdot \boldsymbol{R}_{,\beta}$, 
which encodes local strains, and the coefficients of the second-fundamental 
form, $b_{\alpha\beta}(u,t) =
\hat{\boldsymbol{n}}\cdot\boldsymbol{R}_{,\alpha\beta}$, which describes 
bending, or curvature.  Here, ``$\,\cdot\,$'' is the usual scalar product in 
$\mathbb{R}^3$, a subscript comma followed by an index (say, $\alpha$) is 
shorthand for the partial derivative $\partial/\partial u^\alpha$, and 
$\hat{\boldsymbol{n}}$ is the unit normal to $\mathcal{S}_t$ (see SI).

Assuming no lateral shear ({\it i.e.}, between apical and basal faces) the 
surface of connected basal faces can be written as a normal projection: 
$\boldsymbol{R}_\mathrm{B}(u,t) = \boldsymbol{R}(u,t) - 
\hat{\boldsymbol{n}}(u,t)\ell(u,t)$, where $\ell(u,t)$ is a thickness that can 
vary with both position and time.
%Such a description does not permit lateral shear ({\it i.e.}, between apical 
%and basal faces.
We further assume that the monolayer is ``thin'' in the sense that $\left\vert 
\ell_{,\alpha}\right\vert\ll\left\vert\ell\, H\right\vert \ll 1$, where $H = 
g^{\alpha\beta}\,b_{\alpha\beta}/2$ is the mean curvature of the apical 
surface.  As a result, local geometrical characteristics of the basal surface 
can be expressed as power-series expansions in $\ell$, with coefficients that 
are determined by apical geometry, {\it e.g.},
\begin{equation}
	g_{\alpha\beta}^\mathrm{B}= g_{\alpha\beta} + 2\ell\,b_{\alpha\beta} + 
	\ell^2\left( 2\,H\,b_{\alpha\beta} - K\,g_{\alpha\beta} \right)+
	O\left(\ell^3 \right),
	\label{eq:g_B}
\end{equation}
where $K=b/g$ is the Gaussian curvature [the notation $g=\mathrm{det}\left( 
g_{\alpha\beta} \right)$ and $b=\mathrm{det}\left( b_{\alpha\beta} \right)$ is 
used throughout].  In this approximation, therefore, we need only consider the 
dynamics of a single manifold--- the apical surface $\mathcal{S}_t$--- and the 
field $\ell$.

As the tissue undergoes a deformation, the velocity $\boldsymbol{v}(u,t) = 
\partial\boldsymbol{R}(u,t)/\partial t$ at each point on the apical surface 
causes the coefficients $g_{\alpha\beta}$ and $b_{\alpha\beta}$ to change in 
time (which are, in turn, coupled to $\ell$, as will be shown).  The rate of 
such changes are most naturally expressed in terms of the components of 
$\boldsymbol{v}$ under the decomposition 
$\boldsymbol{v}=v^\alpha\,\boldsymbol{R}_{,\alpha} + 
v^{(n)}\hat{\boldsymbol{n}}$--- {\it i.e.}, tangent- and normal-to 
$\mathcal{S}_t$ (see Fig.~\ref{fig:setup}).  Leaving the details to the SI, we 
have
\begin{equation}
	\partial_t\,g_{\alpha\beta} = v_{\alpha;\beta} + v_{\beta;\alpha}  - 
	2v^{(n)}b_{\alpha\beta},
	\label{eq:par_t_g_comp}
\end{equation}
and
\begin{equation}
	\begin{split}
		\partial_t\,b_{\alpha\beta} &= v^\gamma \,b_{\alpha\beta,\gamma} + 
		b_{\alpha\gamma}\,{v^\gamma}_{,\beta} + 
		b_{\gamma\beta}\,{v^\gamma}_{,\alpha} + v^{(n)}_{,\alpha;\beta}\\
		&\quad- v^{(n)}\left( 2H\,b_{\alpha\beta} - K\,g_{\alpha\beta} \right),
		\label{eq:par_t_b_comp}
	\end{split}
\end{equation}
respectively, where a semi-colon followed by an index is used as shorthand for 
the components of the covariant derivative (see SI).

We make two remarks concerning the above.  First, since not every pair of first 
and second fundamental forms describe a surface, the symmetric, real, 
2$\times$2 matrices of coefficients $g_{\alpha\beta}$ and $b_{\alpha\beta}$ 
only represent four degrees of freedom, rather than six, due to the 
Gauss-Codazzi relations \cite{Frankel}.  Here, such conditions need not be 
explicitly enforced so long as $\mathcal{S}_{t=0}$ is well defined, since 
compatibility with Gauss-Codazzi is preserved under the action of 
Eqs.~(\ref{eq:par_t_g_comp}) and (\ref{eq:par_t_b_comp}).  Second, there is no 
need for an additional equation for the time dependence of the number density 
of cells, $\rho(u,t) = \rho(u,0)/\sqrt{g}$.  By taking the determinant of 
Eq.~(\ref{eq:par_t_g_comp}), followed by the square root, we can deduce that
\begin{equation}
	\partial_t\,\rho + \rho\,{v^\alpha}_{;\alpha} - 2\rho\, v^{(n)} H = 0.
	\label{eq:partial_rho}
\end{equation}
This is precisely the equation for a conserved scalar field associated with a 
moving manifold \cite{Frankel,MarsdenHughes}, but without the standard 
convective term $v^\alpha\,\rho_{,\alpha}$, which does not appear because, by 
construction, there is neither proliferation nor death and cells cannot flow 
relative to the internal coordinate $u$ ({\it cf.} Ref.~\cite{Ranft2010}).

\section{Balance of forces}
The velocity field $\boldsymbol{v}$--- required to close 
Eqs.~(\ref{eq:par_t_g_comp}) and (\ref{eq:par_t_b_comp})--- is prescribed by a 
balance of forces: equating the dissipative forces of the Stokesian embedding 
fluid to those generated in the epithelium.  Ignoring any hydrodynamic affects 
that arise due to finite thickness, the impermeable nature of the tissue 
relates $\boldsymbol{v}$ to the per-unit-area epithelial forces 
$\boldsymbol{f}$ (applied {\it on} the fluid {\it by} the tissue) via 
convolution with the (three dimensional) Oseen tensor 
\cite{HappelBrenner,Peterson1996}
\begin{equation}
	\boldsymbol{v}\left( \boldsymbol{R}\left( u,t \right) \right) = 
	\int_{\mathbb{R}^2}
	%{\left\{u^\prime\in\mathbb{R}^2: \boldsymbol{R}\left( u^\prime,t \right) - 
	%\boldsymbol{R}\left( u,t \right) <C
%\right\}}
\mathrm{d}u^\prime\,\mathsf{O}\left(  \boldsymbol{R}\left( u^\prime,t \right) - 
\boldsymbol{R}\left( u,t \right)\right)\cdot\boldsymbol{f}\left( 
\boldsymbol{R}\left( u^\prime,t \right) \right).
	\label{eq:oseen}
\end{equation}

\begin{figure*}[!t]
\centering
\includegraphics[width=0.9\textwidth]{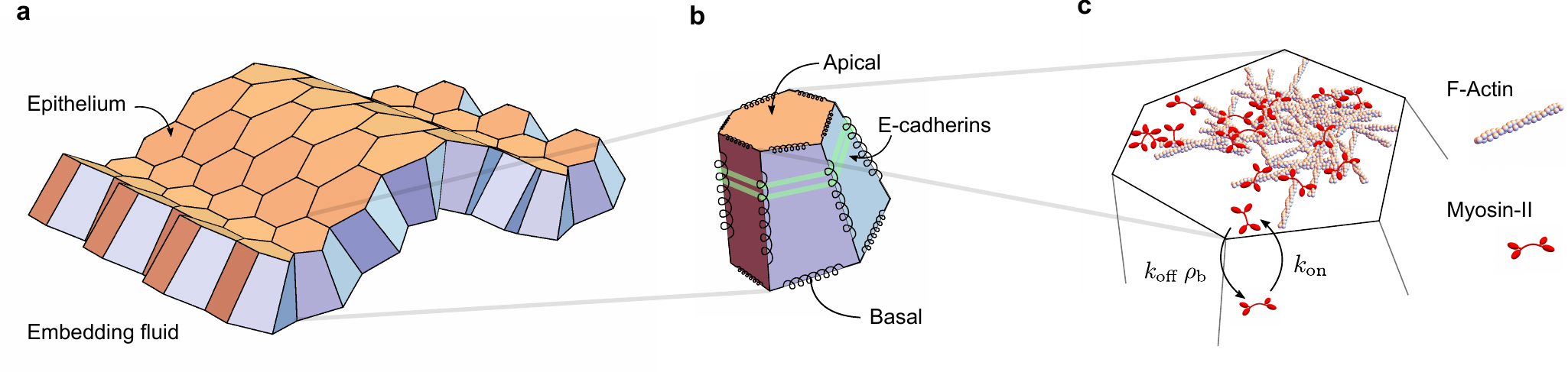}
%/home/rgm/Dropbox/Membranes/Epithelia/Elastomer/Figures/epith_3.pdf
\caption
{
	{\bf Cell-autonomous force generation.} The epithelium is a monolayer of 
	epithelial cells ({\bf a}).  
	%Deformations such as the posterior midgut invagination in Drosophila occur 
	%via an interplay between coordinated force generation, mechanical 
	%response, and dissipation due to the displacement and shear of a low 
	%Reynolds number embedding fluid.
	Each cell is in the shape of a hexagonal prism, with the apical side facing 
	out of the embryo and the basal side facing inwards ({\bf b}).  Cells 
	enclose a fixed (time-independent) volume, and are tightly bound to each 
	other by a localised belt of E-cadherin proteins (cell-cell adhesions) on 
	the lateral faces.  The (active) mechanical response of cells to an applied 
	stress is assumed to be linear [and hence indicated by a spring ({\bf b})], 
	in-line with measurements of recoil following laser ablation.  The action 
	of myosin-II motors on the dense layer of cortical actin that underpins the 
	apical cell faces leads to contractile stresses ({\bf c}).
}
\label{fig:schematic}
\end{figure*}

The forces $\boldsymbol{f}$ can be decomposed into both active and passive-like 
contributions.  Here, passive-like is used as shorthand for dynamical behaviour 
that is characterised by a Lyapounov functional, in analogy with the 
free-energy of a passive system.  We write
\begin{equation}
	\mathcal{F} = \int_{\mathbb{R}^2}F\left( g_{\alpha\beta}, b_{\alpha\beta}, 
	\ell\right)\,\mathrm{vol}^2,
	\label{eq:calF}
\end{equation}
where $\mathrm{vol}^2 = \sqrt{g}\,\mathrm{d}u^1\wedge\mathrm{d}u^2$ is the 
induced volume form on $\mathbb{R}^2$ (see SI) and $F$ is an {\it effective} 
free-energy density (per unit area).  The primary mechanical response of a 
tissue is elastic-like, with restoring forces that are linear in strain (see 
Fig.~\ref{fig:schematic}).  Contributions from apical, basal and lateral faces 
are encoded by five moduli: isotropic and symmetric non-isotropic coefficients 
for each of the apical ($\lambda_\mathrm{A}$ and $\mu_\mathrm{A}$) and basal 
($\lambda_\mathrm{B}$ and $\mu_\mathrm{B}$) sides, and a single coefficient 
$\kappa$ associated with the tissue thickness.  In principle, such quantities 
may rely on the concentrations of passive cross-linkers, cell-cell adhesions or 
other actively-regulated molecules and could even be space-time dependent.  
However, for the purposes of this article they are treated as constant 
(therefore resembling Lam\'{e} coefficients of the first- and second-kind).

Introducing the time-independent matrices $g^\dagger_{\alpha\beta}$ and 
$g^{\mathrm{B}\,\dagger}_{\alpha\beta}$ to represent the reference 
configurations of apical and basal surfaces respectively, the coefficients of 
the corresponding Green-Lagrange strain 2-forms \cite{Frankel,MarsdenHughes} 
are then $\epsilon_{\alpha\beta}  = g_{\alpha\beta} - g^\dagger_{\alpha\beta}$ 
and $\epsilon^\mathrm{B}_{\alpha\beta}  = g_{\alpha\beta} - 
g^{\mathrm{B}\,\dagger}_{\alpha\beta} - 2\,\ell\,b_{\alpha\beta} + 
\ell^2\,\left( 2\,H\,b_{\alpha\beta} - K\,g_{\alpha\beta} \right) + O(\ell^2)$ 
(see SI).  By analogy with both active elastomers 
\cite{Banerjee2011a,Banerjee2017} and discrete (3D) vertex models 
\cite{Hannezo2014,Alt2017,Noll2017a} the effective free-energy is taken to be 
of the form
\begin{equation}
	\begin{split}
	F =& 
	\mu_\mathrm{A}\,\bar{\epsilon}_{\alpha\beta}\,\bar{\epsilon}^{\alpha\beta} 
	+ \lambda_\mathrm{A}\,\left[\mathrm{Tr}_g\left( \epsilon_{\alpha\beta} 
	\right)\right]^2 + 
	\mu_\mathrm{B}\,\bar{\epsilon}^\mathrm{B}_{\alpha\beta}\,\bar{\epsilon}_\mathrm{B}^{\alpha\beta}\\
	& + \lambda_\mathrm{B}\,\left[\mathrm{Tr}_{g_\mathrm{B}}\left( 
	\epsilon^\mathrm{B}_{\alpha\beta} \right)\right]^2
	+\kappa\left( \ell - \ell^\dagger \right)^2,
\end{split}
	\label{eq:Fcompact_2}
\end{equation}
where $\ell^\dagger$ is the reference thickness, and an overbar is used to 
denote the symmetric traceless part, {\it i.e.}, $\bar{\epsilon}_{\alpha\beta} 
= \epsilon_{\alpha\beta} - g^{\alpha\beta}\,\mathrm{Tr}_g \left( 
\epsilon_{\alpha\beta} \right) / 2$\footnotemark[2].
%\footnote{Certain vertex models include an interfacial contribution, 
	%proportional to the total area of the lateral faces 
	%\cite{Hannezo2014,Alt2017}.  However, there is little experimental evidence 
	%to suggest such terms contribute to the energetics at lowest order, 
%therefore they are omitted here for simplicity.}

We assume that the relaxation of the lateral cell faces is orders of magnitude 
faster than either the apical or basal faces, due to a lower density of 
cortical cytoskeleton, implying $\delta \mathcal{F}/\delta \ell = 
0$\footnotemark[3].  The minimisation is performed under the constraint that, 
on the timescales of the midgut invagination, the volume enclosed by epithelial 
cells is incompressible.  That is, the local constraint $V\simeq 
\ell\,\sqrt{g}\left( 1 - \ell\,H \right) + O\left( \ell^3 \right)$,
%
%\begin{equation}
	%V\simeq \ell\,\sqrt{g}\left( 1 - \ell\,H \right) + O\left( \ell^3 \right),
	%\label{eq:V}
%\end{equation}
%
must be satisfied everywhere (see SI) where the $V(u)$ is the time-independent 
volume of a patch containing $\rho(u,0)\,\sqrt{g}(u,0)$ cells.
%As a result, $\ell$ may be eliminated for $g_{\alpha\beta}$ and 
%$b_{\alpha\beta}$ (and material parameters).

%
Using the above to eliminate $\ell$ (in terms of $g_{\alpha\beta}$, 
$b_{\alpha\beta}$, and material parameters), it is the remaining functional 
behaviour of $\mathcal{F}$ that contributes to (\ref{eq:oseen}) via 
\begin{equation}
	\boldsymbol{f} = -\frac{\delta 
		\mathcal{F}}{\delta\boldsymbol{R}}+\nabla\cdot\sigma,
	\label{eq:forces}
\end{equation}
where $\sigma$ is an active stress and $\nabla = 
\boldsymbol{R}_{,\alpha}\,g^{\alpha\beta}\,\partial_\beta$ is the gradient 
operator associated with $\mathcal{S}_t$.  The term 
$\delta\mathcal{F}/\delta\boldsymbol{R}$ may be unpacked (see SI and 
\cite{Guven2004}) to give
\begin{widetext}
	\begin{equation}
		\frac{\delta\mathcal{F}}{\delta\boldsymbol{R}} = \left[ 
			-{\pi^{\alpha\beta}}_{;\beta} - \left( 
			{b^\alpha}_\gamma\,\psi^{\gamma\beta} \right)_{;\beta} - 
			{b^\alpha}_\gamma\,{\psi^{\gamma\beta}}_{;\beta} 
		\right]\,\boldsymbol{R}_{,\alpha} + \left[ 
		-\pi^{\alpha\beta}\,b_{\alpha\beta} - \psi^{\alpha\beta}\left( 
	2H\,b_{\alpha\beta} - K\,g_{\alpha\beta} \right) + 
{\psi^{\alpha\beta}}_{;\alpha\beta} \right]\,\hat{\boldsymbol{n}},
		\label{eq:detaF_deltaR}
	\end{equation}
\end{widetext}
\noindent where
\begin{equation}
	\pi^{\alpha\beta} = \frac{1}{\sqrt{g}}\frac{\partial\left( 
		\sqrt{g}\,F\right)}{\partial\,g_{\alpha\beta}},\ \mathrm{and}\ 
		\psi^{\alpha\beta} =\frac{\partial\,F}{\partial\,b_{\alpha\beta}}.
		\label{eq:pi_varphi}
\end{equation}
In a similar way, the active term may also be expanded to give
\begin{equation}
	\nabla\cdot\sigma = \left({\sigma^{\alpha\beta}}_{;\beta}\right)\, 
	\boldsymbol{R}_{,\alpha} + 
	\left(\sigma^{\alpha\beta}\,b_{\alpha\beta}\right)\,\hat{\boldsymbol{n}}.
	\label{eq:sigma}
\end{equation}
Such forces act at the apical surface only, consistent with experimental 
observations.  If necessary, the effects of active basal stresses can be 
incorporated via ``normal projection'', in the same way as for the effective 
free energy.

\footnotetext[2]{Certain vertex models include an interfacial contribution, 
	proportional to the total area of the lateral faces 
	\cite{Hannezo2014,Alt2017}.  However, there is little experimental evidence 
	to suggest such terms contribute to the energetics at lowest order, 
therefore they are omitted here for simplicity.}
\footnotetext[3]{To incorporate a finite timescale of thickness relaxation, we 
	must use $\partial_t\ell=-\Gamma\,\delta\mathcal{F}/\delta\ell + 
	\mathrm{active\ terms}$.}
For the components of $\sigma$, we make the standard assumption of an unlimited 
background reservoir of ATP, the hydrolysis of which is maintained at a 
constant chemical potential gradient $\Delta \mu_{\mathrm{ATP}}>0$, driving 
contractility.  The stresses that deform the local actin meshwork assume the 
generic form
\begin{equation}
	\sigma^{\alpha\beta} = 
	\chi\left(\rho,\rho_\mathrm{b}\right)\,\Delta\mu_{\mathrm{ATP}}\,g^{\alpha\beta},
	\label{eq:active}
\end{equation}
where $\chi$ is a compressibility that not only relies on the density of cells, 
$\rho$, but also the density of apically bound myosin-II, $\rho_\mathrm{b}$ 
(see Fig.~\ref{fig:schematic}).  The scalar field $\rho_\mathrm{b}$ is 
non-conserved and drives tangent-plane contractility at the apical surface.  As 
a first approximation, we assume simple Langmuir-like kinetics, where myosin-II 
filaments bind to apical actin from the bulk of a cell with a rate 
$k_{\mathrm{on}}$, and unbind at a rate which is proportional to the current 
density $k_{\mathrm{off}}\rho_b$.  The resulting continuity equation is 
therefore modified to
\begin{equation}
	\partial_t \rho_b + \left( \rho_b \,v^{\alpha} \right)_{;\alpha} - 
	\rho_b\,v^{(n)}\,2H =
	%D\Delta\rho_b k_b\,\rho
	k_{\mathrm{on}} - k_{\mathrm{off}}\rho_b.
	%e^{ \mathsf{A}:\mathsf{\epsilon}},
	\label{eq:CoMy}
\end{equation}

\begin{figure*}[t]
\centering
\includegraphics[width=0.98\textwidth]{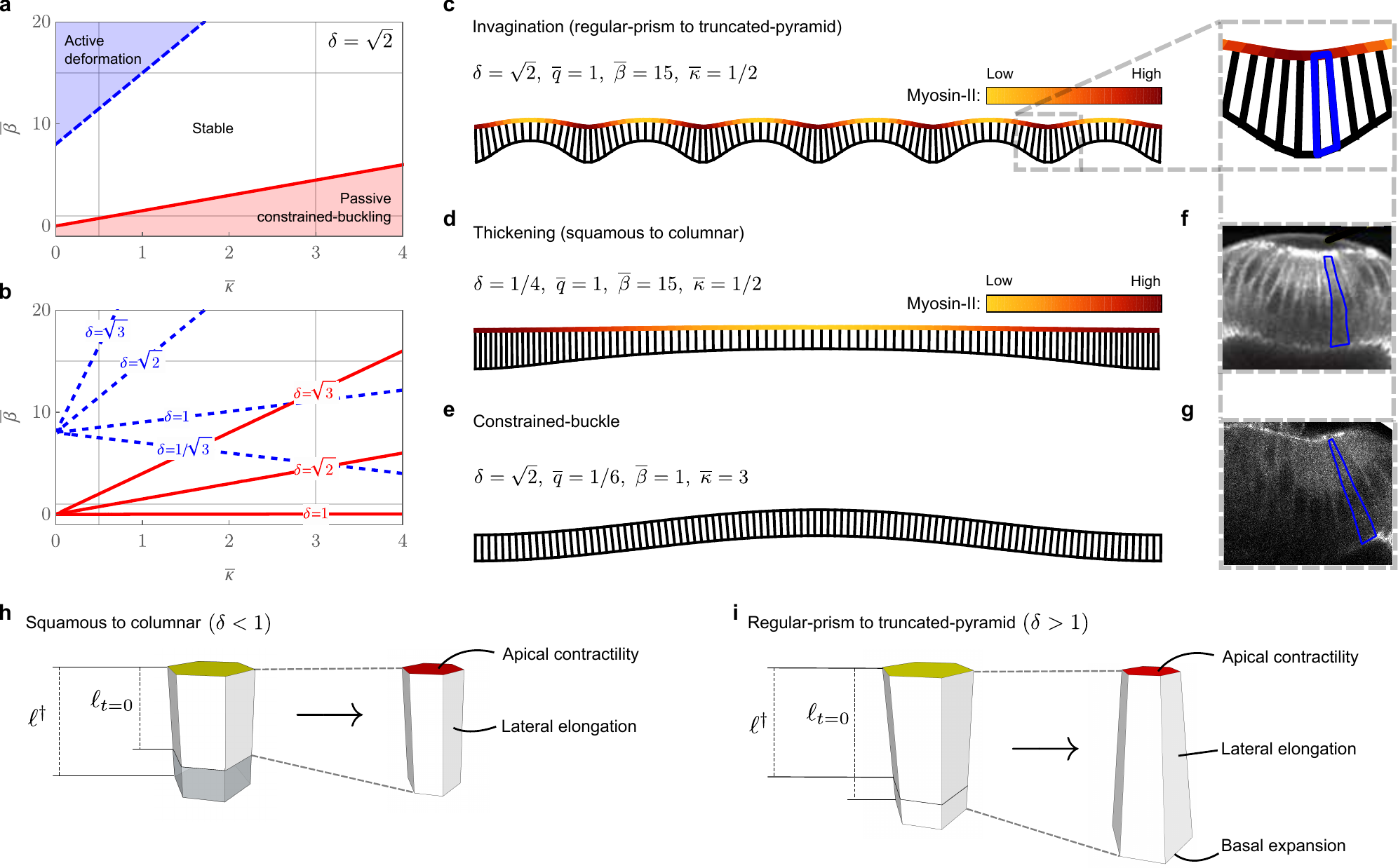}
%/home/rgm/Dropbox/Membranes/Epithelia/Elastomer/Figures/phases_2.pdf
\caption
{
	{\bf Hydrodynamic instabilities reproduce the key features observed during 
	the onset of invaginations.} Two classes of instability can be identified 
	({\bf a}): passive constrained-buckling and active deformation.  The latter 
	can be further categorised into two sub-types: invaginating and thickening 
	[({\bf c}) and ({\bf d}), respectively].  Boundaries between stable and 
	un-stable configurations are controlled by $\delta = \ell_{t=0} / 
	\ell^\dagger$, with the constrained-buckling instability disappearing for 
	all $\delta \le 1$ ({\bf b}).  Geometrically faithful schematics of active 
	[({\bf a}) and ({\bf b})] and passive ({\bf c}) instabilities can be 
	constructed (see SI).  For the former, the concentration of apical 
	myosin-II is shown via a colormap, ranging from yellow (low concentration) 
	to red (high concentration).  In terms of cellular shape changes, the 
	invaginating and thickening instabilities are characterised by 
	``regular-prism to truncated-pyramid'' ({\bf i}) and ``squamous to 
	columnar'' ({\bf h}) transitions, respectively.  The invaginating 
	instability ({\bf c}) qualitatively reproduces the key features of both 
	mesoderm ({\bf f}) and midgut ({\bf g}) invaginations, observed using 
	multi-photon microscopy of regulatory light-chain-GFP labelled myosin.  
	Increased myosin density [shown in white in ({\bf f}) and ({\bf g})] at the 
	apical surface correlates with constriction of the cell faces resulting in 
	both lateral and basal extension.  The former thickens the Epithelium 
	whilst the latter causes splay and hence induces curvature.  Image ({\bf 
	f}) was reproduced from \cite{plosone} with permission, whilst image ({\bf 
	g}) was kindly contributed by T.~Lecuit, C.~Collinet and A.~Bailles (see 
Acknowledgements).}
\label{fig:phases}
\end{figure*}

\section{Hydrodynamic Instabilities}
The system of equations (\ref{eq:par_t_g_comp}), (\ref{eq:par_t_b_comp}) and 
(\ref{eq:CoMy}) describe the dynamical behaviour of five degrees-of-freedom 
(the remaining equations are constitutive, describing $\boldsymbol{v}$ as a 
function of $\rho_\mathrm{b}$, $g_{\alpha\beta}$ and $b_{\alpha\beta}$).  The 
treatment is fully-covariant and capable of describing arbitrary tissue 
deformations so long as they do not change the topology of the apical manifold 
({\it e.g.} by introducing holes or handles).  Due to their inherent non-linear 
nature, however, the closed system of PDEs we present cannot be solved 
analytically.  Nevertheless, the system is, in principle, amenable to numerical 
implementation via the use of coordinate-free finite-element methods 
\cite{CompFluidMech,HigherOrder}.  Such an approach also allows the study (and 
visualisation) of complex geometries, permitting much greater comparison with 
experiment.  Leaving this task for further work, the remainder of this article 
is dedicated to the study of characteristic behaviours by employing linear 
stability analysis.

For simplicity, the analysis is restricted to deformations that are confined to 
a single plane (so-called ``quasi-1D''), reducing the number of dynamical 
degrees-of-freedom to three--- {\it e.g.}, $\sqrt{g}$, $H$ and 
$\rho_\mathrm{b}$.  In this simplified setup, we consider the stability (in 
Fourier space \cite{IntegralTransforms}) of perturbations from a flat, up/down 
symmetric, steady state at time $t=0$ [{\it i.e.}, $b_{\alpha\beta}(u,0) = 0$ 
for all $u$].  For convenience, we assume that the rest configurations of 
apical and basal surfaces are equal 
($g^{\mathrm{B}\,\dagger}_{\alpha\beta}=g^\dagger_{\alpha\beta}$) and that the 
Lam\'{e}-like coefficients are also equal 
($\lambda_\mathrm{A}=\lambda_\mathrm{B}=:\lambda$ and 
$\mu_\mathrm{A}=\mu_\mathrm{B}=:\mu$).  This ensures that no torque need be 
applied at the tissue boundary in order to keep it flat.  We do, however, apply 
an in-plane force at the boundary.  The reason is that, although flat, the 
tissue is active: there is a homogeneous steady-state concentration of bound 
myosin-II $\rho_\mathrm{b}(u,0)$, which leads to homogeneous contractile 
stresses.  We choose the boundary force to be such that the (homogeneous) 
metric at steady state, $g_{\alpha\beta}(u,0)$, is the same as the rest metric 
$g^\dagger_{\alpha\beta}$.  The coordinates $u$ may then be chosen such that 
$g_{\alpha\beta}(u,0) = g^\dagger_{\alpha\beta} = \delta_{\alpha\beta}$.

Consigning the details to the SI, the results are shown in 
Fig.~\ref{fig:phases}.  Here, instabilities are given in terms of a 
dimensionless measure of active contractility,  
$\overline{\beta}:=\chi^{(0)}\,\Delta\mu_{\mathrm{ATP}} / \left( \lambda + 
\mu/2 \right)$, and a dimensionless (elastic-like) response due to changes in 
thickness, $\overline{\kappa}:=\kappa\,\left( \ell^\dagger \right)^2 / \left( 
\lambda + \mu/2 \right)$.  The boundary between stable and unstable regions is 
characterised by a dimensionless ``pre-strain'' $\delta = \ell(u,0) / 
\ell^\dagger$--- {\it i.e.}, the steady state ($t=0$) strain of the lateral 
faces.  Given the implied boundary force described above, this is tantamount to 
choosing the volume enclosed by each cell.
%(and hence the pressure difference between the cell and the embedding fluid). 

For comparatively low $\overline{\beta}/\overline{\kappa}$, the tissue is 
unstable to passive {\it constrained}-buckling (below the solid red line of 
Fig.~\ref{fig:phases}a).  The instability only exists for $\delta>1$--- {\it 
i.e.}, when the lateral faces are extended at steady state--- and is contrary 
to predictions of traditional un-constrained buckling \cite{Hannezo2014}, based 
on energy minimisation arguments.  Constrained-buckling (Fig.~\ref{fig:phases}e 
and SI) is a {\it finite} wave-number effect seen, for example, in poroelastic 
rods \cite{Skotheim2004} and microtubules in living cells 
\cite{Brangwynne2006a}.  Here, the characteristic wave-number of the buckle is 
non-zero due to the combined effects of basal and lateral elasticity.  The 
former imposes an effective bending energy $\sim\delta\left( \lambda + \mu/2 
\right)$, suppressing basal expansion and therefore buckling at large 
wave-number.  For the latter, the low curvature deformations at small 
wave-number require lateral extension, with an energy cost $\sim\kappa$.

For comparatively high $\overline{\beta}/\overline{\kappa}$, {\it active} 
deformations are unstable (above the dashed blue line of 
Fig.~\ref{fig:phases}a) and correspond to a contractile instability at the 
apical surface, driven by variations in $\rho_\mathrm{b}$.  However, the effect 
that this has on cell shapes, and hence the overall tissue deformation, depends 
on $\delta$.

For $\delta\le 1$, the lateral faces are under compression, and cells respond 
to apical contraction by elongation--- {\it i.e.}, a squamous to columnar 
transition.  Here, the shape of the apical surface remains unchanged, whilst 
the thickness is inversely proportional to apical contractility, vanishing as 
wave-number approaches zero (Fig.~\ref{fig:phases}d and SI). 

By contrast, if $\delta> 1$, the lateral faces are already under strain.  As a 
result, myosin-driven apical constriction leads not only to lateral extension, 
but also to an expansion of the basal face--- {\it i.e.}, a regular-prism to 
truncated-pyramid transition.  The corresponding tissue-scale deformations 
(Fig.~\ref{fig:phases}c and SI) qualitatively resemble early-onset 
invaginations seen during Drosophila gastrulation (Figs.~\ref{fig:phases}f and 
\ref{fig:phases}g).  This result is also clearly in-line with {\it in vivo} 
cellular tomography, where two-photon scanning microscopy images of 
E-cadherin-GFP mutants permits the high-fidelity three-dimensional 
reconstruction of cell shapes \cite{Gelbart2012}.  Here, apical constriction of 
cells at the mesoderm invagination results in cell shapes that are extended 
both laterally and basally, whilst keeping cell volume constant.         
%Here, apical constriction leads not only to thickening of the monolayer (via 
%lateral extension) but also curvature generated by cellular ``splay'' ({\it 
%i.e.}, basal extension).

\section{Discussion}
In summary, we use a thin-film approximation and differential geometry to 
capture the salient physical features of epithelial monolayer morphology: 
apical contractility; mechanical response; and a momentum-conserving embedding 
fluid (Fig.~\ref{fig:schematic}).  While our general treatment allows us to 
deal with tissues of arbitrary curvature, analysis in a restricted scenario 
identifies three distinct hydrodynamic instabilities that are qualitatively 
in-line with experimental evidence.  We envisage such regimes to be common 
across a broad range of deforming tissues.

Importantly, our approach sets the foundations for further refinements in the 
understanding of tissue physics.  Many outstanding questions can be addressed 
within the context of the framework set out here, for example: what happens in 
more complex geometries ({\it e.g.}, in systems where the steady-state symmetry 
is broken); how is the mechanical response of a tissue actively regulated ({\it 
e.g.}, how do $\kappa$ and the four Lam\'{e}-like coefficients rely on 
$\rho_\mathrm{b}$); what is the role, if any, of active interfacial dynamics 
between lateral faces ({\it e.g.}, \cite{Bielmeier2016a}); how do cells 
modulate their behaviour in response to mechanical cues, such as stress, 
strain, or strain-rate ({\it e.g.}, how do the rates $k_\mathrm{on}$ and 
$k_\mathrm{off}$ rely on $\epsilon_{\alpha\beta}$, $\pi^{\alpha\beta}$ {\it 
etc}.), and; what is the role of the embedding fluid, and other global 
constraints such as the surrounding inextensible vitelline membrane. 

Experimentally, advances in {\it in vivo} imaging 
\cite{Pantazis2014,Mavrakis2010a} combined with classical genetics and 
state-of-the-art mechano-biology manipulations ({\it e.g.}, laser-ablation and 
laser-induced cauterisation) are leading renewed interest in the behaviour of 
tissues during development \cite{Keller2012,Heisenberg2013}, suggesting that a 
data-led approach to such questions will soon be within reach.  In this 
context, our closed-form dynamical theory provides a quantitative framework 
which, due to its covariant nature, can be used in conjunction with data 
concerning shape changes such as movement, compression, folding, and 
invagination.  That is, over and above the presented linear stability analysis, 
we expect a numerical implementation of the full non-linear theory to prove a 
useful tool for understanding the increasingly sophisticated experiments of 
developmental biology.

\section{Acknowledgements}
We thank Simons Foundation (USA) for financial support.  For detailed 
discussion, and the contribution of the image used in Fig.~4g, we thank 
T.~Lecuit, C.~Collinet and A.~Bailles from IBDM, Universit\'{e} Aix-Marseille.  
We acknowledge the CNRS (Laboratoire International Associ\'{e} between IBDM and 
NCBS) for associated travel support.  We also thank S.~Gadgil (IISc), 
K.~Husain, A.~Rautu and A.~Singh (all NCBS) for helpful discussions.

\vspace{5mm}
\section{Author Contributions}
RGM and MR conceived of the project and wrote the manuscript together.  RGM 
performed research and analysis, with guidance from MR.

\section{Competing Financial Interests}
The authors declare no conflict of interest.

%\bibliographystyle{INSERT}
%\bibliography{/home/rgm/Documents/Papers/Bibtex/library}

\clearpage

\begin{widetext}
	\section{Supplementary Material}

\section{Introduction}\label{sec:intro}
This supplementary material comprises a set of calculations (including a terse 
summary of the necessary background and notation), the details of which would 
be needed in order to recapitulate the results presented in the main 
manuscript. Useful references regarding technical aspects are 
\cite{Frankel,MarsdenHughes}. 

\section{Setup}\label{sec:setup}
Consider a time-dependent manifold $\mathcal{S}_t$, comprising points 
$\boldsymbol{R}(u,t) =: \boldsymbol{R}_t(u)$ that are the image of a Lagrangian 
coordinate $u\in\mathbb{R}^2$ (with components $u^\alpha$, for $\alpha=1,2$) 
under the embedding, $\boldsymbol{R}_t:\mathbb{R}^2\to \mathbb{R}^3$.  Let 
$\vec{e}_\alpha$ be basis vectors spanning $T_u\mathbb{R}^2$ ({\it i.e.}, the 
tangent space to $\mathbb{R}^2$ at point $u$), and $\hat{\boldsymbol{e}}_i$ be 
the usual Euclidean basis spanning $T_{\boldsymbol{R}(u,t)}\mathbb{R}^3$ [{\it 
i.e.}, the tangent space to $\mathbb{R}^3$ at point $\boldsymbol{R}(u,t)$, 
which is just $\mathbb{R}^3$ with the origin translated by 
$\boldsymbol{R}(u,t)$].  Throughout, we adopt the convention that Greek indices 
take values 1 or 2, whilst Latin indices take 1, 2, or 3.  If 
$\boldsymbol{R}_{t\,\ast}:T_u\mathbb{R}^2\to 
T_{\boldsymbol{R}(u,t)}\mathbb{R}^3$ is the pushforward of $\boldsymbol{R}_t$, 
then $\boldsymbol{R}_{t\,\ast} \left(\vec{e}_\alpha\right) =\partial 
\boldsymbol{R}(u,t) / \partial u^\alpha =: \boldsymbol{R}_{,\alpha}$, where a 
subscript comma followed by an index (say, $\alpha$) is shorthand for the 
partial derivative $\partial/\partial u^\alpha$.  Using this notation 
$\hat{\boldsymbol{n}} = \left(\boldsymbol{R}_{,1} \times 
\boldsymbol{R}_{,2}\right) / \left\vert \boldsymbol{R}_{,1}\times 
\boldsymbol{R}_{,2}\right\vert$ is the unit normal to $\mathcal{S}_t$.  The 
basis of 1-forms $\mathrm{d}u^\alpha$ ($\alpha=1,2$) span 
$T^\ast_u\mathbb{R}^2$ and are dual to the $\vec{e}_\alpha$, such that 
$\mathrm{d}u^\alpha (\vec{e}_\beta) = \delta^\alpha_\beta$, where 
$\delta^\alpha_\beta$ is the Kronecker delta symbol.

%It should be remarked that, under a change of coordinates, the transformation 
%properties of the functions $v^i$ (see \S \ref{sec:not}) and $p_i$ can be 
%readily calculated.  The former is said to transform in a {\it contravariant} 
%way, and the latter in a {\it covariant} way.

\subsection{Metric, raising and lowering}\label{sec:met}
The embedding function $\boldsymbol{R}_t$ induces a metric on $\mathbb{R}^2$ 
via the pullback $\boldsymbol{R}_t^\ast$.  That is $g_{\alpha\beta}(u) = \left( 
\boldsymbol{R}_t^\ast\,\mathrm{I} \right)\left( \vec{e}_\alpha,\vec{e}_\beta 
\right) = \boldsymbol{R}_{t\,\ast}\left(\vec{e}_\alpha 
\right)\cdot\boldsymbol{R}_{t\,\ast}\left(\vec{e}_\beta \right) 
=\boldsymbol{R}_{,\alpha} \cdot \boldsymbol{R}_{,\beta}$, where $\mathrm{I}$ is 
the first fundamental form of $\mathbb{R}^3$ ({\it i.e.}, with coefficients 
$\delta_{ij}$) and ``$\cdot$'' is the usual dot product of $\mathbb{R}^3$.  At 
each point $u\in\mathbb{R}^2$, the induced metric can be used to define an 
inner product $\langle \cdot,\cdot\rangle:T_u\mathbb{R}^2\times 
T_u\mathbb{R}^2\to \mathbb{R}$.  That is, for arbitrary vectors $\vec{v}$ and 
$\vec{w}$, we define $\langle\vec{v},\vec{w}\rangle :=
v^\alpha g_{\alpha\beta} w^\beta$.  Such an inner product permits the explicit 
identification of a vector, {\it e.g.}, $\vec{v}$, with its dual 1-form, $v$, 
by the condition $v(\vec{w})=\langle \vec{v},\vec{w}\rangle$, which holds for 
all $\vec{w}$.  Noticing that $v(\vec{w}) = v_\alpha \mathrm{d}u^\alpha 
(\vec{w}) = v_\alpha w^\alpha$ and using the above definition of the inner 
product of two vectors implies the raising and lowering properties of the 
metric and its inverse [$g^{\alpha\beta} = \left( g_{\alpha\beta} 
\right)^{-1}$], respectively.  That is, $v_\alpha = g_{\alpha\beta} v^\beta$ 
and $v^\alpha = g^{\alpha\beta} v_\beta$.  Using this property, the inner 
product acting on two 1-forms can be defined in a complementary way to that of 
the inner product on vectors:
\begin{equation}
	\langle v,w\rangle := v_\alpha g^{\alpha\beta} w_\beta = v_\alpha w^\alpha= 
	\langle\vec{v},\vec{w}\rangle.
	\label{eq:inner_1_form}
\end{equation}

\subsection{Volume form and wedge product}
The induced (pseudo)-volume-$2$-form on $\mathbb{R}^2$ is given by  
\begin{equation}
	\mathrm{vol}^2 = \sqrt{g}\,\mathrm{d}u^1\wedge\mathrm{d}u^2,
	\label{eq:vol^2}
\end{equation}
where the shorthand $g = \det{\left[g_{\alpha\beta}\right]}$ has been used.  
Here, the symbol ``$\wedge$'' indicates a {\it wedge} product.  If $p$ and $q$ 
are 1-forms, their wedge product is given by $p\wedge q = p\otimes q - q\otimes 
p$, from which it is clear that $p\wedge q = -q\wedge p$.  More generally, the 
wedge product is bilinear and associative, and has the following commutation 
relation: if $p^{(r)}$ and $q^{(s)}$ are forms of degree $r$ and $s$, 
respectively, then $p^{(r)} \wedge q^{(s)} = \left( -1 \right)^{r\,s}\,q^{(s)} 
\wedge p^{(r)}$.  The space of all $k$-forms created by taking wedge products 
of 1-forms is written as $\bigwedge^k T^\ast\mathcal{S}_t$.  Such forms are 
{\it alternating} linear functionals.  (Note: 0-forms and 1-forms are 
considered to be alternating forms, even though it is ill-defined to ask if 
they are skew-symmetric).

\subsection{Covariant derivative}\label{sec:covar}
The covariant derivative represents the rate of change of a tensor field (at 
$u$) whilst moving along the unique geodesic that has tangent vector with 
pre-image $\vec{y}$ (at $u$) under $\boldsymbol{R}_{t\,\ast}$.  In our setup:

\begin{itemize}
	\item When acting on a scalar field $\phi$, we write
\begin{equation}
	\nabla_{\vec{y}} \,\phi := \phi_{,\alpha}\,\mathrm{d}u^\alpha (\vec{y}),
	\label{eq:cov_scalar}
\end{equation}
where a subscript comma ``,'' is shorthand for a partial derivative, {\it 
i.e.}, ${\phi}_{,\alpha} := \partial \phi / \partial u^\alpha$.

\item When acting on a vector $\vec{v} = v^\alpha \vec{e}_\alpha$, we write
\begin{equation}
	\nabla_{\vec{y}} \,\vec{v} := \vec{e}_\alpha \left( 
	{v^\alpha}_{;\beta}\right)\mathrm{d}u^\beta (\vec{y}),
	\label{eq:cov_vec}
\end{equation}
where the components ${v^\alpha}_{;\beta}$ are given by 
\begin{equation}
	{v^\alpha}_{;\beta} := {v^\alpha}_{,\beta} + v^\gamma 
	\Gamma^\alpha_{\beta\gamma}.
	\label{eq:cov_components}
\end{equation}
Once again, a subscript comma ``,'' is shorthand for a partial derivative, 
${v^i}_{,j} := \partial v^i / \partial u^j$, whilst the 
$\Gamma^\alpha_{\beta\gamma} = g^{\alpha\delta}\left(g_{\delta\beta,\gamma} + 
g_{\delta\gamma,\beta} - g_{\beta\gamma,\delta}\right)/2$ are Christoffel 
symbols, which define the action of the covariant derivative, via 
$\nabla_{\vec{e}_\alpha} \vec{e}_\beta = \vec{e}_\gamma 
\Gamma^\gamma_{\alpha\beta}$.  Note that the shorthand $\nabla_\alpha := 
\nabla_{\vec{e}_\alpha}$ is frequently used in physics.

\item For a $1$-form, the action of the covariant derivative can be defined by 
	demanding that the ``Leibniz rule'' holds.  That is, if a scalar field is 
	defined by the action of a $1$-form on a vector, {\it i.e.}, 
	$\phi:=v(\vec{w})=v^\alpha\,w_\alpha$, then
\begin{equation}
	\nabla_\alpha \left(v^\beta\,w_\beta\right) = 
	\left(v^\beta\,w_\beta\right)_\alpha := {v^\beta}_{;\alpha}\, w_\beta + 
	v^\beta \,w_{\beta;\alpha}.
	\label{eq:Leibnitz}
\end{equation}
The result is that $v_{\alpha;\beta}:= v_{\alpha,\beta} - 
v_\gamma\,\Gamma^\gamma_{\alpha\beta}$, 
%
%\begin{equation}
	%v_{\alpha;\beta}:= v_{\alpha,\beta} - 
	%v_\gamma\,\Gamma^\gamma_{\alpha\beta},
	%\label{eq:cov_1form}
%\end{equation}
%
which is consistent with the notion of using the metric as a raising / lowering 
operator ({\it i.e.}, $v_{\alpha;\beta}=g_{\alpha\gamma}{v^\gamma}_{;\beta}$).  
In coordinate free notation, this is equivalent to
\begin{equation}
	\left(\nabla_{\vec{y}} \,v \right) \left( \vec{w} \right) := 
	\nabla_{\vec{y}} \left[v\left( \vec{w} \right)\right] - v\left( 
	\nabla_{\vec{y}} \,\vec{w} \right).
	\label{eq:cov_1_form}
\end{equation}

\item For a general $(r,s)$-valent tensor $T:\left(\otimes^r\, 
	T\mathcal{S}_t\right)\otimes\left(\otimes^s\, 
	T^\ast\mathcal{S}_t\right)\to\mathbb{R}$, acting on $r$ vectors 
	$\vec{w}_1,\dots,\vec{w}_r\in T\mathcal{S}_t$ and $s$ 1-forms 
	$p_1,\dots,p_s\in T^\ast\mathcal{S}_t$, the covariant derivative is given 
	by the following formula:
\begin{equation}
	\begin{split}
	\left( \nabla_{\vec{y}}\,T \right)\left( \vec{v}_1, \ldots, \vec{v}_r, 
	p_1,\ldots,p_s \right) :=& \,\vec{y}\left[ T \left( \vec{v}_1, \ldots, 
	\vec{v}_r, p_1,\ldots,p_s \right)\right] \\
	&- T\left(\nabla_{\vec{y}}\,\vec{v}_1, \ldots, \vec{v}_r, p_1,\ldots,p_s 
	\right) - \ldots - T\left(\vec{v}_1, \ldots, \nabla_{\vec{y}}\,\vec{v}_r, 
	p_1,\ldots,p_s \right)\\
	&-T\left( \vec{v}_1, \ldots, \vec{v}_r, \nabla_{\vec{y}}\,p_1,\ldots,p_s 
	\right) - \ldots - T\left(\vec{v}_1, \ldots, \vec{v}_r, 
	p_1,\ldots,\nabla_{\vec{y}}\,p_s \right).
\end{split}
	\label{eq:cov_T}
\end{equation}
In component form, this is equivalent to
\begin{equation}
	\begin{split}
	\nabla_{\vec{y}}\, T :=& \bigg( y^\gamma 
	{T^{\beta_1\ldots\beta_s}}_{\alpha_1\ldots\alpha_r,\gamma}
	-{T^{\beta_1\ldots\beta_s}}_{\delta\ldots\alpha_r}\Gamma^\delta_{\alpha_1\gamma}\,y^\gamma 
	- \ldots - 
	{T^{\beta_1\ldots\beta_s}}_{\alpha_1\ldots\delta}\Gamma^\delta_{\alpha_r\gamma}\,y^\gamma\\
	&+{T^{\delta\ldots\beta_s}}_{\alpha_1\ldots\alpha_r}\Gamma^{\beta_1}_{\delta\gamma}\,y^\gamma 
	+ \ldots + 
	{T^{\beta_1\ldots\delta}}_{\alpha_1\ldots\alpha_r}\Gamma^{\beta_s}_{\delta\gamma}\,y^\gamma\bigg)\, 
	\mathrm{d}u^{\alpha_1}\otimes \ldots\otimes\mathrm{d}u^{\alpha_r}\otimes 
	\vec{e}_{\beta_1}\otimes\ldots\otimes\vec{e}_{\beta_s}.
\end{split}
	\label{eq:cov_T_comp}
\end{equation}
\end{itemize}

\subsection{Second fundamental form, Gauss and Weingarten 
equations}\label{sec:II}
Consider the derivative
\begin{equation}
	%^{\mathbb{R}^3}\nabla_{\boldsymbol{F}_\ast(\vec{e}_\alpha)} 
	%\,\hat{\boldsymbol{n}} = \frac{\partial \hat{\boldsymbol{n}}}{\partial 
	%x^i}\frac{\partial x^i}{\partial u^\alpha}=
	\frac{\partial\hat{\boldsymbol{n}}}{\partial 
	u^\alpha}=:\hat{\boldsymbol{n}}_{,\alpha},	\label{eq:dn_dualpha}
\end{equation}
{\it i.e.}, the rate-of-change in the unit normal to $\mathcal{S}_t$ along 
$u^\alpha$, expressed as a vector in $\mathbb{R}^3$.  Since 
$\hat{\boldsymbol{n}}$ is a unit vector, the result must still be tangent to 
$\mathcal{S}_t$ and therefore
\begin{equation}
	\hat{\boldsymbol{n}}_{,\alpha} = -{b^\beta}_\alpha 
	\,\boldsymbol{R}_{t\,\ast}(\vec{e}_\beta) = -{b^\beta}_\alpha 
	\,\boldsymbol{R}_{,\beta},
	\label{eq:Wiengarten}
\end{equation}
which is known as the Weingarten equation (the assignment of a minus sign being 
convention). Given the right-hand side, we can use the coefficients from the 
above to construct a linear map $b:T_u\mathbb{R}^2\to T_u\mathbb{R}^2$ by 
writing $\vec{b}(\vec{v})=-v^\beta \vec{e}_\alpha\, {b^\alpha}_\beta$, for 
arbitrary $\vec{v}$.  Similarly, there is a natural bilinear form 
$\mathrm{II}$, know as the {\it second fundamental form}, that can be 
associated with such a map, whose action is given by
\begin{equation}
	\mathrm{II}\left(\vec{v},\vec{w}\right) = 
	\left\langle\vec{v},\vec{b}(\vec{w})\right\rangle =  v^\gamma 
	w^\beta\left\langle\vec{e}_\gamma,-{b^\alpha}_\beta\, \vec{e}_\alpha 
	\right\rangle.
	\label{eq:B}
\end{equation}
That is
\begin{equation}
	\mathrm{II}
	%= -\left\langle 
	%\mathrm{d}\boldsymbol{R},\mathrm{d}\hat{\boldsymbol{n}}\right\rangle
	= - \left[\boldsymbol{R}_{t\,\ast} 
	\left(\vec{e}_\alpha\right)\cdot\hat{\boldsymbol{n}}_{,\beta}\right]\,\mathrm{d}u^\alpha\otimes\mathrm{d}u^\beta 
	= - 
	\left[\boldsymbol{R}_{,\alpha}\cdot\hat{\boldsymbol{n}}_{,\beta}\right]\,\mathrm{d}u^\alpha\otimes\mathrm{d}u^\beta=b_{\alpha\beta}\,\mathrm{d}u^\alpha\otimes\mathrm{d}u^\beta,
	\label{eq:b_ij}
\end{equation}
where $b_{\alpha\beta}=g_{\alpha\gamma}b^\gamma_\beta$.  Notice that since 
$\partial \left[\boldsymbol{R}_{,\alpha}\cdot\hat{\boldsymbol{n}}\right] / 
\partial u^\beta = 0$, we have $b_{\alpha\beta} =  \left(\partial 
\boldsymbol{R}_{,\alpha}/\partial u^\beta\right)\cdot \hat{\boldsymbol{n}} = 
\boldsymbol{R}_{,\alpha\beta}\cdot \hat{\boldsymbol{n}}$.  More generally, the 
derivative of basis vectors $\vec{e}_\alpha$ with respect to some coordinate 
$u^\beta$ on $\mathcal{S}_t$ can be decomposed into tangent and normal parts.
%
%\begin{equation}
	%\frac{\partial \boldsymbol{F}_\ast(\vec{e}_{\alpha})}{\partial u^\beta} = 
	%\Gamma^{\gamma}_{\alpha\beta}\boldsymbol{F}_\ast(\vec{e}_\gamma) + 
	%b_{\alpha\beta}\,\hat{\boldsymbol{n}},
	%\label{eq:Gauss}
%\end{equation}
%
\begin{equation}
	\boldsymbol{R}_{,\alpha\beta} = 
	\Gamma^\gamma_{\alpha\beta}\,\boldsymbol{R}_{,\gamma} + 
	b_{\alpha\beta}\,\hat{\boldsymbol{n}},
	\label{eq:Gauss}
\end{equation}
which is known as Gauss' equation.

\subsection{Curvature}\label{sec:curv}
At a given point $u$, each unit vector $\hat{\vec{y}}$ corresponds to a unique 
curve $\mathcal{C}$ on $\mathcal{S}_t$ that also lies in the plane 
$\mathcal{P}_t$ spanned by $\hat{\boldsymbol{n}}$ and 
$\boldsymbol{R}_{t\,\ast}\left(\hat{\vec{y}}\right)=y^\alpha\,\boldsymbol{R}_{,\alpha}$.  
The action of the second fundamental form on a given $\hat{\vec{y}}$ results in 
the {\it normal} curvature $c^{(n)}_{\hat{\vec{y}}}$ of $\mathcal{S}_t$ in the 
direction of $y^\alpha\,\boldsymbol{R}_{,\alpha}$ ({\it i.e.}, the curvature of 
$\mathcal{C}$ in $\mathcal{P}_t$).  We write,
\begin{equation}
	\mathrm{II}(\hat{\vec{y}},\hat{\vec{y}}) = \pm c^{(n)}_{\hat{\vec{y}}},
	\label{eq:kappa}
\end{equation}
where ``+'' indicates whether $\mathcal{C}$ is curving towards the unit normal, 
and vice-versa for ``-''.  Since the normal curvature will change dependent on 
which direction $\hat{\vec{y}}$ is chosen, we define the principal directions:
\begin{equation}
	\hat{\vec{y}}_1 (p) = \argmax_{\hat{\vec{y}}\in 
	T_p\mathbb{R}^2}\,\mathrm{II}(\hat{\vec{y}}, \hat{\vec{y}}),\ \mathrm{and}\ 
	\hat{\vec{y}}_2(p) = \argmin_{\hat{\vec{y}}\in T_p\mathbb{R}^2}\, 
	\mathrm{II}(\hat{\vec{y}}, \hat{\vec{y}}).
	\label{eq:T_12}
\end{equation}
The principal curvatures are then given by
\begin{equation}
	c_\alpha(p) = \mathrm{II}(\hat{\vec{y}}_\alpha,\hat{\vec{y}}_\alpha),\ 
	\forall\,  \alpha = 1,2.
	\label{eq:kappa_12}
\end{equation}
It can be shown that the $c_\alpha$ are eigenvalues of the linear operator 
$\vec{b}$ from \S \ref{sec:II}.  That is,
\begin{equation}
	\vec{b}(\hat{\vec{y}}_\alpha) = c_\alpha\,\hat{\vec{y}}_\alpha,
	\label{eq:princ}
\end{equation}
where if $c_1\neq c_2$, the principal directions are orthogonal.  We may now 
define two different types of local curvature of $\mathcal{S}_t$: the {\it 
mean} curvature 
\begin{equation}
	H := \frac{1}{2}\mathrm{Tr}\,{b^\alpha}_\beta = 
	\frac{1}{2}\mathrm{Tr}_g\,\mathrm{II} = \frac{c_1+c_2}{2},
	\label{eq:H}
\end{equation}
and the {\it Gaussian} curvature
\begin{equation}
	K:=\mathrm{det}\,{b^\alpha}_\beta = 
	\frac{\mathrm{det}\,b_{\alpha\beta}}{\mathrm{det}\,g_{\alpha\beta}} = 
	c_1\,c_2.
	\label{eq:K}
\end{equation}

\subsection{Interior product and trace}\label{app:int}
The action of a differential form on a vector is essentially a contraction.  
(To see this, consider a 1-form $v$ acting on a vector $\vec{w}$: $v(\vec{w}) = 
v^\alpha w_\alpha$).  For forms of higher degree, if $p^{(r)}$ is an $r$-form 
and $\vec{v}$ a vector, then $i_{\vec{v}} p^{(r)}$ is an $(r-1)$-form--- {\it 
i.e.}, the {\it interior} product of $p^{(r)}$ with respect to $\vec{v}$.  If, 
$\vec{w}_1, \vec{w}_2, \dots,\vec{w}_r$ are $r$ arbitrary vectors, then
\begin{equation}
	i_{\vec{v}} p^{(r)}\left(\vec{w}_1, \dots,\vec{w}_{r-1}\right) = 
	p^{(r)}\left(\vec{v}, \vec{w}_1, \dots,\vec{w}_{r-1}\right).
	\label{eq:interior}
\end{equation}
The interior product $i_{\vec{v}}:\bigwedge^k T^\ast\mathcal{S}_t 
\to\bigwedge^{k-1} T^\ast\mathcal{S}_t$ is an {\it antiderivation}, which means 
that its action over the wedge product is given by 
\begin{equation}
	i_{\vec{v}}\left[ p^{(r)}\wedge q^{(s)} \right]=\left[ i_{\vec{v}} p^{(r)} 
	\right] \wedge q^{(s)} + (-1)^r p^{(r)} \wedge \left[i_{\vec{v}} q^{(s)} 
	\right].
	\label{eq:antideriv}
\end{equation}
The interior product contracts a form with a vector, reducing the degree of the 
form.  For contractions between {\it pairs} of coefficients of a form, the 
notion of a generalised trace $\mathrm{Tr}_g^{(a,b)}$ is needed.  Here, the 
subscript $g$ indicates a trace {\it through} the metric ({\it i.e.}, to ensure 
that contractions only take place between indices of different types) whilst 
the integers $a$ and $b$ indicate the indices over which to contract.  For 
example, the trace of $r$-form $p^{(r)}$ gives a $(r-2)$-form:
\begin{equation}
	\mathrm{Tr}^{(1,3)}_g\left[ p^{(r)} \right]\left(\vec{w}_1, 
	\dots,\vec{w}_{r-2}\right) = p^{(r)}\left(\vec{e}_i, \vec{w}_1, \vec{e}_j, 
	\dots,\vec{w}_{r-2}\right)\,g^{ij}.
	\label{eq:Tr_form}
\end{equation}

\subsection{Exterior derivative}\label{app:d}
The exterior derivative $\mathrm{d}:\bigwedge^k T^\ast\mathcal{S}_t 
\to\bigwedge^{k+1} T^\ast\mathcal{S}_t$ takes $k$-forms and returns 
$(k+1)$-forms.  In our setup:

\begin{itemize}
	\item If $\varphi$ is a 0-form, or function, then
\begin{equation}
	\mathrm{d}\varphi = \left( \frac{\partial \varphi}{\partial u^1} 
	\right)\mathrm{d}u^1 + \left( \frac{\partial \varphi}{\partial u^2} 
	\right)\mathrm{d}u^2.
	\label{eq:df}
\end{equation}
By the identification of vectors with 1-forms (see \S \ref{sec:met}) 
$\mathrm{d}\varphi$ is dual to the gradient of $\varphi$.  That is, 
$\mathrm{d}\varphi(\vec{v}) = \langle \nabla \varphi,\vec{v}\rangle$ for all 
vectors $\vec{v}$, where
\begin{equation}
	\nabla \varphi := g^{\alpha\beta}\frac{\partial \varphi}{\partial u^\alpha} 
	\vec{e}_\beta,
	\label{eq:grad}
\end{equation}
is just the gradient operator induced by the embedding.
	\item If $\omega = \omega_\alpha \mathrm{d}u^\alpha$ is a 1-form, then
\begin{equation}
	\mathrm{d}\omega=\left[ \left( \frac{\partial \omega_2}{\partial u^1} 
	\right) - \left( \frac{\partial \omega_1}{\partial u^2} \right) 
\right]\mathrm{d}u^1\wedge\mathrm{d}u^2.
	\label{eq:dalpha^0}
\end{equation}
\end{itemize}
Repeated application always yields zero--- {\it i.e.}, 
$\mathrm{d}\mathrm{d}p=\mathrm{d}^2 p=0$ for an arbitrary exterior form $p$.  
Also, the action of $\mathrm{d}$ is an {\it antiderivation}.  That is, 
\begin{equation}
	\mathrm{d}\left[p^{(r)}\wedge q^{(s)}\right] = \mathrm{d}p^{(q)}\wedge 
	q^{(s)} + (-1)^r p^{(r)}\wedge \mathrm{d}q^{(s)},
	\label{eq:dantideriv}
\end{equation}
where $p^{(r)}$ and $q^{(s)}$ are exterior forms of degree $r$ and $s$, 
respectively.

\section{Morphology}\label{sec:morph}
We assume that the surface of connected Apical faces may be approximated by the 
manifold $S_t$.  For $S_t$ to change in time, each point $\boldsymbol{R}(u,t) 
=: \boldsymbol{R}_t(u)$ must move with a velocity $\boldsymbol{v}(u,t) 
:=\partial_t\boldsymbol{R}(u,t) \in T_{\boldsymbol{R}(u,t)}\mathbb{R}^3$.  In 
the following, we show how the local structure of $\mathcal{S}_t$ changes as 
points move under the action of $\boldsymbol{v}$. 

\subsection{Rate-of-change of the metric in time}\label{sec:rate_met}
We wish to calculate the partial derivative $\partial_t g_{\alpha\beta}$.  
Applying the product rule, we have
\begin{equation}
	\partial_t g_{\alpha\beta} = \left(\partial_t 
	\boldsymbol{R}_{,\alpha}\right) \cdot \boldsymbol{R}_{,\beta} + 
	\boldsymbol{R}_{,\alpha} \cdot \left(\partial_t 
	\boldsymbol{R}_{,\beta}\right).
	\label{eq:product}
\end{equation}
Here, since the coordinates $u$ do not depend on time, the partial derivates 
$\partial /\partial t$ and $\partial / \partial u^\alpha$ commute, giving
\begin{equation}
	\partial_t g_{\alpha\beta} = \boldsymbol{v}_{,\alpha} \cdot 
	\boldsymbol{R}_{,\beta} + \boldsymbol{R}_{,\alpha} \cdot 
	\boldsymbol{v}_{,\beta}.
	\label{eq:commute}
\end{equation}
Decomposing $\boldsymbol{v}$ into parts that are tangent- and normal-to 
$\mathcal{S}_t$, we have $\boldsymbol{v} = v^\alpha\,\boldsymbol{R}_{,\alpha} + 
v^{(n)}\hat{\boldsymbol{n}}$, which can be substituted into 
$\boldsymbol{v}_{,\alpha}$ to give
\begin{equation}
	\boldsymbol{v}_{,\alpha} = {v^\beta}_{,\alpha}\,\boldsymbol{R}_{,\beta} + 
	v^\beta\,\boldsymbol{R}_{,\beta\alpha} + 
	{v^{(n)}}_{,\alpha}\,\hat{\boldsymbol{n}} + 
	v^{(n)}\,\hat{\boldsymbol{n}}_{,\alpha}.
	\label{eq:v_alpha}
\end{equation}
Imposing the Gauss (\ref{eq:Gauss}) and Wiengarten (\ref{eq:Wiengarten}) 
equations, we see that
\begin{equation}
	\boldsymbol{v}_{,\alpha} = \left( {v^\beta}_{,\alpha} + 
	v^\gamma\,\Gamma^\beta_{\alpha\gamma} - v^{(n)}\,{b_\alpha}^\beta 
	\right)\boldsymbol{R}_{,\beta} + \left( b_{\alpha\beta}\,v^\beta + 
	{v^{(n)}}_{,\alpha} \right)\hat{\boldsymbol{n}}.
	\label{eq:v_alpha_Gauss_Weingarten}
\end{equation}
Substituting (\ref{eq:v_alpha_Gauss_Weingarten}) into (\ref{eq:commute}), the 
raising and lowering properties of the metric and the relation 
$g_{\alpha\beta,\gamma} = 
\Gamma_{\alpha\gamma}^\kappa\,g_{\kappa\beta}+\Gamma_{\beta\gamma}^\kappa\,g_{\kappa\alpha}$ 
may be used to show that
\begin{equation}
	\partial_t\,g_{\alpha\beta} = v_{\alpha;\beta} + v_{\beta;\alpha} - 
	2\,v^{n)}\,b_{\alpha\beta},
	\label{eq:dt_g_result}
\end{equation}
where a subscript colon ``;'' is used to denote the components of the covariant 
derivative (see \S \ref{sec:covar}).
%
%\begin{equation}
	%v_{\alpha;\beta} = v_{\alpha,\beta} - 
	%v_\gamma\,\Gamma^\gamma_{\alpha\beta}.
	%\label{eq:covar_1_form}
%\end{equation}

\subsection{Rate-of-change of local area in time}\label{sec:rate_area}
The local area at a point $u$ on $\mathcal{S}_t$ is just given by $\sqrt{g}$, 
where
\begin{equation}
	g:=\mathrm{det}\,g_{\alpha\beta} = 
	\frac{1}{2}\varepsilon^{\alpha\mu}\,\varepsilon^{\beta\nu}\,g_{\alpha\beta}\,g_{\mu\nu},
	\label{eq:detg}
\end{equation}
for 2$\times$2 symmetric matrices.  Here, $\varepsilon^{\alpha\beta}$ is a 
two-dimensional antisymmetric Levi-Civita symbol (a rank-$(0,2)$ tensor density 
of weight $+1$).  Taking the partial derivative with respect to time yields
\begin{equation}
	\partial_t\,\sqrt{g}=\frac{1}{2\sqrt{g}}
	\varepsilon^{\alpha\mu}\,\varepsilon^{\beta\nu}\,g_{\alpha\beta}
	\left(\partial_t\, g_{\mu\nu}\right) = \frac{1}{2\sqrt{g}}
	\varepsilon^{\alpha\mu}\,\varepsilon^{\beta\nu}\,g_{\alpha\beta}
	\left(v_{\alpha;\beta} + v_{\beta;\alpha} - 
	2\,v^{n)}\,b_{\alpha\beta}\right)
	\label{eq:d_t_area}
\end{equation}
where (\ref{eq:dt_g_result}) has been used.  Using the fact that
\begin{equation}
	g^{\alpha\beta} = \frac{1}{g} 
	\varepsilon^{\alpha\mu}\,\varepsilon^{\beta\nu} g_{\mu\nu},
	\label{eq:inverse}
\end{equation}
gives the result
\begin{equation}
	\partial_t\,\sqrt{g}=\sqrt{g}\left( {v^\alpha}_{;\alpha} - v^{(n)}\,2H 
	\right).
	\label{eq:d_area_dt_result}
\end{equation}

\subsection{Relationship to conservation of cell number density}\label{sec:rho}
Dividing (\ref{eq:d_area_dt_result}) by $\sqrt{g}$ we see that
\begin{equation}
	\partial_t\,\log\left( \frac{\sqrt{g}}{\sqrt{g_0}} \right) = 
	{v^\alpha}_{;\alpha} - v^{(n)}\,2H,
	\label{eq:log_sqrt_g}
\end{equation}
where a subscript is used for notational convenience [{\it i.e.}, 
$g_0(u)=g(u,0)$].  However, since there is no cell division or death, the 
tissue is of fixed connectivity and cannot flow relative to $u$, implying 
$\rho_0\,\sqrt{g_0}=\rho\,\sqrt{g}$, where $\rho(u,t)$ is the local number 
density (per unit area) of cells.  As a result
\begin{equation}
	\log\left( \frac{\sqrt{g}}{\sqrt{g_0}} \right) = -\log\left( 
	\frac{\rho}{\rho_0} \right),
	\label{eq:log_rho}
\end{equation}
and hence
\begin{equation}
	\partial_t\,\rho + \rho\,{v^\alpha}_{;\alpha} - \rho\, v^{(n)}\,2 H = 0.
	\label{eq:partial_rho_SI}
\end{equation}
This is precisely the equation for a conserved scalar field associated with a 
moving manifold, but without the standard convective term 
$v^\alpha\,\rho_{;\alpha}$, which does not appear because, by construction, 
cells cannot flow relative to the internal coordinate $u$.  [Note that this can 
	also be seen as a consequence of the fact that $\sqrt{g}$ is a rank-0 
	tensor density of weight +1, and therefore its covariant derivative 
vanishes, {\it i.e.}, $\left(\sqrt{g}\right)_{;\alpha} = 0$].

\subsection{Rate-of-change of the coefficients of the second fundamental form 
in time}\label{sec:rate_b}
In a similar way as for the metric, consider calculating $\partial_t
\,b_{\alpha\beta}$, where $b_{\alpha\beta} = 
\boldsymbol{R}_{,\alpha\beta}\cdot\hat{\boldsymbol{n}}$ are the coefficients of 
the second fundamental form.  Using the product rule, we have
\begin{equation}
	\partial_t\,b_{\alpha\beta} = \left(\partial_t\, 
	\boldsymbol{R}_{,\alpha\beta}\right)\cdot\hat{\boldsymbol{n}} + 
	\boldsymbol{R}_{,\alpha\beta}\cdot\left( \partial_t\, 
	\hat{\boldsymbol{n}}\right),
	\label{eq:dt_b}
\end{equation}
where we consider each of the two terms on the right-hand side in turn.  For 
the first term, by commuting time and space derivatives as before, we see that 
$\partial_t\, \boldsymbol{R}_{,\alpha\beta} = \boldsymbol{v}_{,\alpha\beta}$.  
Taking the partial derivative $\partial/\partial u^\beta$ of 
(\ref{eq:v_alpha}), applying Gauss' equation (\ref{eq:Gauss}), and retaining 
only components in the $\hat{\boldsymbol{n}}$ direction, gives
\begin{equation}
	\left(\partial_t\, 
	\boldsymbol{R}_{,\alpha\beta}\right)\cdot\hat{\boldsymbol{n}} = 
	\boldsymbol{v}_{,\alpha\beta}\cdot\hat{\boldsymbol{n}}=\left( 
	{v^{\gamma}}_{;\alpha} - v^{(n)}\,{b_\alpha}^\gamma\right)b_{\gamma\beta} + 
	\left( b_{\alpha\gamma}\,v^\gamma \right)_{,\beta} + 
	v^{(n)}_{,\alpha\beta}.
	\label{eq:v_alphabeta}
\end{equation}
For the second term, it is necessary to understand 
$\partial_t\,\hat{\boldsymbol{n}}$.  Here, since $\partial_t \vert 
\hat{\boldsymbol{n}}\vert^2 = 0$, we see that 
$\hat{\boldsymbol{n}}\cdot\left(\partial_t\,\hat{\boldsymbol{n}}\right)=0$, 
{\it i.e.}, $\partial_t\,\hat{\boldsymbol{n}}$ has no normal component.  
Moreover, since $\boldsymbol{R}_{,\alpha}\cdot\hat{\boldsymbol{n}} = 0$ then 
$\boldsymbol{v}_{,\alpha}\cdot \hat{\boldsymbol{n}} = 
\boldsymbol{R}_{,\alpha}\cdot\left(\partial_t\,\hat{\boldsymbol{n}}\right)$, 
which implies
\begin{equation}
	\partial_t\,\hat{\boldsymbol{n}} = 
	-v^\beta\,{b_\beta}^\alpha\,\boldsymbol{R}_{,\alpha} - 
	v^{(n)}_{,\alpha}\,g^{\alpha\beta}\,\boldsymbol{R}_{,\beta}.
	\label{eqq:dt_n}
\end{equation}
Substituting into $\boldsymbol{R}_{,\alpha\beta}\cdot\left(\partial_t 
\,\hat{\boldsymbol{n}}\right)$ and once again using Gauss' equation 
(\ref{eq:Gauss}), gives
\begin{equation}
	\boldsymbol{R}_{,\alpha\beta}\cdot\left(\partial_t 
	\,\hat{\boldsymbol{n}}\right) = 
	\Gamma^\gamma_{\alpha\beta}\,v^\kappa\,b_{\kappa\gamma} - 
	\Gamma^\gamma_{\alpha\beta}\,v^{(n)}_{,\gamma}.
	\label{eq:second_terms}
\end{equation}
The expressions (\ref{eq:v_alphabeta}) and (\ref{eq:second_terms}) can be 
substituted into (\ref{eq:dt_b}) with the following results:
\begin{align}
	%\begin{split}
	\partial_t\,b_{\alpha\beta} &=\ b_{\gamma\beta}\,{v^\gamma}_{;\alpha} + 
	\left( b_{\alpha\beta}\,v^\beta \right)_{;\alpha} + v^{(n)}_{,\alpha;\beta} 
	- v^{(n)}\,{b_\alpha}^\gamma\,b_{\gamma\beta}\\
	&= b_{\gamma\beta}\,{v^\gamma}_{;\alpha} + 
	b_{\alpha\gamma}\,{v^\gamma}_{;\beta} + v^\gamma\,b_{\alpha\beta;\gamma} + 
	v^{(n)}_{,\alpha;\beta} - v^{(n)}\left( 2H\,b_{\alpha\beta} - 
	K\,g_{\alpha\beta} \right).
%\end{split}
	\label{eq:dt_b_final}
\end{align}
Here, the last line uses the relation ${b_\alpha}^\gamma\,b_{\gamma\beta} = 
2H\,b_{\alpha\beta} - K\,g_{\alpha\beta}$, and the fact that the coefficients 
of the covariant derivative of a rank-(2,0) tensor (\S\ref{sec:covar}) are 
\begin{equation}
	b_{\alpha\beta;\gamma} = b_{\alpha\beta,\gamma} - 
	b_{\alpha\kappa}\,\Gamma^\kappa_{\beta\gamma} - 
	b_{\kappa\beta}\,\Gamma^\kappa_{\alpha\gamma}.
	\label{eq:cov_der_ten}
\end{equation}

\subsection{Rate-of-change of mean curvature in time}
Using the above results, it is possible to calculate the rate of change of mean 
curvature (\ref{eq:H}) in time.  We have
\begin{equation}
	\partial_t\,H = \frac{1}{2}\left( \partial_t\, b_{\alpha\beta} 
	\right)g^{\alpha\beta} + \frac{1}{2}b_{\alpha\beta}\left( 
	\partial_t\,g_{\alpha\beta} \right) = \frac{1}{2}\left( \partial_t\, 
	b_{\alpha\beta} \right)g^{\alpha\beta} - 
	\frac{1}{2}b^{\alpha\beta}\left(\partial_t\,g_{\alpha\beta} \right),
	\label{eq:d_H_dt}
\end{equation}
where, in the last equality we use the fact that $\partial_t\, 
\delta^\alpha_\beta =0$, which implies
\begin{equation}
	\partial_t\,g^{\alpha\beta} = 
	-g^{\alpha\mu}\,g^{\beta\nu}\,\partial_t\left( g_{\mu\nu} \right).
	\label{eq:inverse_trick}
\end{equation}
Substituting for the results (\ref{eq:dt_g_result}), (\ref{eq:dt_b_final}) and
using $b_{\alpha\beta}b^{\alpha\beta} = 4H^2 - 2K$, gives
\begin{equation}
	\partial_t\,H = v^\alpha\,H_{,\alpha} + \frac{1}{2}\Delta v^{(n)} + 
	v^{(n)}\left( 2H^2 - K \right),
	\label{eq:d_H_dt_result}
\end{equation}
where $\Delta := \partial_\alpha\left( 
\sqrt{g}\,g^{\alpha\beta}\,\partial_\beta \right)/\sqrt{g}$ is the 
Laplace-Beltrami operator.

\section{Mechanical response}\label{sec:free-energy}
The mechanical response of the tissue under deformation is assumed to be 
elastic-like, with restoring forces that are linear in strain corresponding to 
each of the apical, basal and lateral cell faces.

\subsection{Apical}\label{sec:passive_apical}
Consider first the Apical manifold $\mathcal{S}_t$.  Using the theory of finite 
strains, we consider the difference between two configurations: an unstrained 
state $\mathcal{S}^\dagger$ and the state at time $t$, $\mathcal{S}_t$.  To do 
this, we introduce $\boldsymbol{\phi}:\mathbb{R}^2\to\mathbb{R}^3$, the 
embedding function for the unstrained state $\mathcal{S}^\dagger$ [{\it i.e.}, 
$\boldsymbol{\phi}(u,t)=\boldsymbol{R}^\dagger(u,t)$ and 
$\boldsymbol{\phi}_\ast(\vec{e}_\alpha) = \boldsymbol{R}^\dagger_{,\alpha}$].  
The coefficients of the pullback (to $\mathbb{R}^2$) of the Green-Lagrange 
strain form are then
\begin{equation}
	\epsilon_{\alpha\beta} = \left[ \left( \boldsymbol{R}_t^\ast - 
	\boldsymbol{\phi}^\ast \right) \mathrm{I} \right]\left( \vec{e}_\alpha, 
	\vec{e}_\beta \right) = 
	\boldsymbol{R}_{,\alpha}\cdot\boldsymbol{R}_{,\beta} - 
	\boldsymbol{R}^\dagger_{,\alpha}\cdot\boldsymbol{R}^\dagger_{,\beta}= 
	g_{\alpha\beta} - g^\dagger_{\alpha\beta},
	\label{eq:strain}
\end{equation}
where $\mathrm{I} = \delta_{ij}\,\mathrm{d}x^i\otimes\mathrm{d}x^j$ is just the 
first fundamental form of three-dimensional Euclidean space, and 
$g^\dagger_{\alpha\beta}$ is the metric of the unstrained state.

We then construct an effective free energy density (per unit area) for the 
Apical faces that is quadratic in the coefficients $\epsilon_{\alpha\beta}$.  
The most general way is to contract two copies of (\ref{eq:strain}) with a 
rank-(0,4) elasticity tensor {\it i.e.}, 
$C^{\alpha\beta\gamma\delta}\,\epsilon_{\alpha\beta}\,\epsilon_{\gamma\delta}$.  
For our treatment, it suffices to follow the usual decomposition of $C$ between 
trace and symmetric-traceless parts: 
\begin{equation}
	C^{\alpha\beta\gamma\delta} = \mu_{\mathrm{A}}\left( 
	g^{\alpha\gamma}g^{\beta\delta} - 
	\frac{1}{2}g^{\alpha\beta}g^{\gamma\delta} \right) + \left( 
	\lambda_\mathrm{A} + \frac{\mu_{\mathrm{A}}}{2} 
	\right)g^{\alpha\beta}g^{\gamma\delta},
	\label{eq:Lame}
\end{equation}
such that $\mu_{\mathrm{A}}$ and $\lambda_{\mathrm{A}}$ resemble the first and 
second Lam\'{e} coefficients of the Apical surface, respectively.  (Note that, 
in principle, the values of $\mu_{\mathrm{A}}$ and $\lambda_{\mathrm{A}}$ can 
rely on the concentrations of wide variety of molecules, from passive 
cross-linkers to cell-cell adhesions {\it etc}.).

\subsection{Basal}
The surface of connected Basal faces is also assumed to be approximated by a 
smooth Riemannian manifold, $\mathcal{S}^\mathrm{B}_t$.  In a ``thin film'' 
approximation, we make the assumption that $\mathcal{S}^\mathrm{B}_t$ is given 
by by taking a normal projection from $\mathcal{S}_t$, such that 
$\boldsymbol{R}_\mathrm{B} (u,t) = \boldsymbol{R} (u,t) - 
\ell(u,t)\,\hat{\boldsymbol{n}}(u,t)$, where $\ell$ is the tissue thickness.  
The metric of the Basal manifold can then be written in terms of the metric of 
the Apical manifold via power series expansion in $\left\vert 
\ell\,H\right\vert \ll 1$, {\it e.g.},
\begin{equation}
	g^\mathrm{B}_{\alpha\beta} = g_{\alpha\beta} + 2\ell\,b_{\alpha\beta} + 
	\ell^2\left( 2\,H\,b_{\alpha\beta} - K\,g_{\alpha\beta} \right) + 
	\ell_{,\alpha}\ell_{,\beta} +
	O\left(\ell^3 \right),
	\label{eq:g_basal}
\end{equation}
and
\begin{equation}
	g_\mathrm{B}^{\alpha\beta} = g^{\alpha\beta} - 2\ell\,b^{\alpha\beta} + 3 
	\ell^2\left( 2\,H\,b_{\alpha\beta} - K\,g_{\alpha\beta} \right) - 
	\ell_{,\alpha}\ell_{,\beta} +
	O\left(\ell^3 \right).
	\label{eq:g_basal_inverse}
\end{equation}
As explained in the main manuscript, we neglect the final terms in these 
expansions, making the assumption that $\left\vert \ell_{,\alpha}\right\vert 
\ll \ell\,\left\vert H\right\vert$.
The effective free-energy density (per unit area) of the basal surface is then 
$C_\mathrm{B}^{\alpha\beta\gamma\delta}\,\epsilon_{\alpha\beta}^\mathrm{B}\,\epsilon_{\gamma\delta}^\mathrm{B}$, 
where the coefficients of the basal Green-Lagrange strain form 
(\S\ref{sec:passive_apical}) are
\begin{equation}
	\epsilon_{\alpha\beta}^{\mathrm{B}} = g_{\alpha\beta}^{\mathrm{B}} - 
	g^{\mathrm{B}\,\dagger}_{\alpha\beta},
	\label{eq:basal_strain}
\end{equation}
and
\begin{equation}
	C^{\alpha\beta\gamma\delta}_\mathrm{B} = \mu_\mathrm{B}\left( 
	g^{\alpha\gamma}_\mathrm{B}\,g^{\beta\delta}_\mathrm{B} - 
	\frac{1}{2}\,g^{\alpha\beta}_\mathrm{B}\,g^{\gamma\delta}_\mathrm{B} 
	\right) + \left( \lambda_\mathrm{B} + \frac{\mu_{\mathrm{B}}}{2} 
	\right)g^{\alpha\beta}_\mathrm{B}\,g^{\gamma\delta}_\mathrm{B},
	\label{eq:Basal_lame}
\end{equation}
such that $\mu_{\mathrm{B}}$ and $\lambda_{\mathrm{B}}$ may differ from  
$\mu_{\mathrm{A}}$ and $\lambda_{\mathrm{A}}$.

\subsection{Lateral}
The lateral cell walls are assumed to have a rest length $\ell^\dagger$ and 
therefore have a contribution to the effective free energy of the form 
$\kappa\left( \ell - \ell^\dagger \right)^2$, where $\kappa$ has the dimensions 
of an elastic modulus.  Some vertex models have included an interfacial 
contribution, proportional to the total area of lateral faces.  However, there 
is scant experimental evidence that such terms contribute at to energetics at 
lowest order, therefore they are omitted for simplicity.

\subsection{Volume constraint}
Summing contributions from apical, basal and lateral faces, the total effective 
free-energy of the tissue is given by
\begin{equation}
	\mathcal{F}=\int_{\mathbb{R}^2} F\left( \boldsymbol{R}_{,\alpha}, 
	\boldsymbol{R}_{;\alpha\beta}, \ell\right)\,\mathrm{vol}^2.
	\label{eq:sriptF_ell}
\end{equation}
Since the density of cortical actin associated with the lateral faces is lower 
than that of either the apical or basal faces, we assume that the relaxation 
time of $\ell$ is significantly faster than $g_{\alpha\beta}$, and hence we 
impose $\delta\mathcal{F}/\delta\ell = 0$.  The minimisation must be performed 
under the constraint that cells do not change their volume.  Once again 
appealing to the thin film approximation, the area of lateral slice of tissue, 
projected a distance $z$ in the $-\hat{\boldsymbol{n}}$ direction from 
$\mathcal{S}_t$ is $\sqrt{g_z} = \sqrt{g}\left( 1 + 2\,z\,H \right) + 
O\left(z^2\right)$, and therefore
\begin{equation}
	V=\int_0^\ell\,\mathrm{d}z\,\sqrt{g_z} = \ell\,\sqrt{g}\left( 1 + \ell\,H 
	\right) + O\left( \ell^3 \right),
	\label{eq:V_constraint}
\end{equation}
where $V$ is time-independent.

\subsection{Variation of the effective free-energy}
Using the aforementioned procedure to eliminate $\ell$ results in an effective 
free-energy functional whose dynamical degrees of freedom are purely 
geometrical, {\it i.e.},
\begin{equation}
	\mathcal{F}=\int_{\mathbb{R}^2} F\left( \boldsymbol{R}_{,\alpha}, 
	\boldsymbol{R}_{;\alpha\beta} \right)\,\mathrm{vol}^2,
	\label{eq:sriptF}
\end{equation}
where $\mathrm{vol}^2 = \sqrt{g}\,\mathrm{d}u^1\wedge\mathrm{d}u^2$ is the 
induced volume form on $\mathbb{R}^2$.  We wish to calculate the functional 
derivative of $\mathcal{F}$ with respect to variations in the embedding 
$\boldsymbol{R}(u,t)$.  That is, the response of the system ({\it i.e.}, the 
forces per unit area) to deformations of $\mathcal{S}_t$.

Formally, we consider the variation $\delta\mathcal{F}$ arising due to changes 
to $\mathcal{S}_t$ of the form $\boldsymbol{R} + \epsilon\,\boldsymbol{\eta}$, 
where $\epsilon$ is a small dimensionless parameter.  We write  
\begin{equation}
	\delta\mathcal{F} = 
	\epsilon\,\left.\frac{d\mathcal{F}}{d\epsilon}\right\vert_{\epsilon=0} = 
	%\epsilon\int_{\mathbb{R}^2}\boldsymbol{\eta}\cdot\left( 
	%\frac{\delta\mathcal{F}}{\delta\boldsymbol{R}} \right)=
	\epsilon\int_{\mathbb{R}^2} \left( 
	\boldsymbol{\eta}_{,\alpha}\cdot\boldsymbol{A}^{\alpha} + 
	\boldsymbol{\eta}_{;\alpha\beta}\cdot\boldsymbol{B}^{\alpha\beta} 
	\right)\mathrm{vol}^2,
	\label{eq:deltaF_1}
\end{equation}
where
\begin{equation}
	\boldsymbol{A}^{\alpha} := \frac{1}{\sqrt{g}}\frac{\partial\left( 
		\sqrt{g}\,F \right)}{\partial\boldsymbol{R}_{,\alpha}}
		%= \frac{1}{\sqrt{g}}\left(\frac{\partial\left( \sqrt{g}\,F 
	%\right)}{\partial\,g_{\alpha\beta}}\right)^{\alpha\beta}\boldsymbol{R}_{,\beta} 
	= \pi^{\alpha\beta}\boldsymbol{R}_{,\beta},
	\label{eq:A^alpha}
\end{equation}
and
\begin{equation}
	\boldsymbol{B}^{\alpha\beta}
	:=\frac{\partial\,F}{\partial\,\boldsymbol{R}_{;\alpha\beta}}=
	%\Lambda^{\alpha\beta\gamma}\,\boldsymbol{R}_{,\gamma} + 
	\psi^{\alpha\beta}\hat{\boldsymbol{n}},
	\label{eq:B^alphabeta}
\end{equation}
By using the chain rule and the fact that 
$\partial\,g/\partial\,g_{\alpha\beta} = g\,g^{\alpha\beta}$ [see 
(\ref{eq:detg}) and (\ref{eq:inverse})], it can be shown that
\begin{equation}
	\pi^{\alpha\beta} = \frac{1}{\sqrt{g}}\frac{\partial\left( 
		\sqrt{g}\,F\right)}{\partial\,g_{\alpha\beta}} = 
		\frac{\,F\,g^{\alpha\beta}}{2} + 
		\frac{\partial\,F}{\partial\,g_{\alpha\beta}}.\label{eq:pi}
\end{equation}
Similarly, from Gauss' equation (\ref{eq:Gauss}) and orthogonality 
($\boldsymbol{R}_{,\mu}\cdot\hat{\boldsymbol{n}} = 0$) we have

%it is clear that $\boldsymbol{R}_{,\alpha\beta} = 
%b_{\alpha\beta}\,\hat{\boldsymbol{n}}$, and hence
%
\begin{equation}
	%\Lambda^{\alpha\beta\gamma} = 
	%\frac{\partial\,F}{\partial\,\Gamma^\gamma_{\alpha\beta}},\ \mathrm{and}\ 
	\psi^{\alpha\beta} =\frac{\partial\,F}{\partial\,b_{\alpha\beta}}.
	\label{eq:Lambda_varphi}
\end{equation}
%
%and
%%
%\begin{equation}
	%\varphi^{\alpha\beta} =\frac{\partial\,F}{\partial\,b_{\alpha\beta}}.
	%\label{eq:varphi}
%\end{equation}
%
We proceed by integrating by parts which, in the language of differential 
geometry, is just an application of Stokes theorem \cite{Frankel}. For the 
first term on the right-hand side of (\ref{eq:deltaF_1}) let $\omega$ be a 
1-form, such that
\begin{equation}
	\omega = 
	\boldsymbol{\eta}\cdot\boldsymbol{R}_{,\alpha}\,\pi^{\alpha\beta}\,i_{\vec{e}_\beta}\mathrm{vol}^2,
	\label{eq:omega}
\end{equation}
where $i$ represents the interior product (\S \ref{app:int}).  Applying the 
exterior derivative (\S \ref{app:d}), we have
\begin{equation}
	\mathrm{d}\omega = 
	\boldsymbol{\eta}_{,\beta}\cdot\boldsymbol{R}_{,\alpha}\,\pi^{\alpha\beta}\,\mathrm{vol}^2
	+ 
	\boldsymbol{\eta}\cdot\left(\boldsymbol{R}_{,\alpha}\,{\pi^{\alpha\beta}}_{;\beta} 
	+ 
	\hat{\boldsymbol{n}}\,\pi^{\alpha\beta}\,b_{\alpha\beta}\right)\mathrm{vol}^2,
	\label{eq:domega}
\end{equation}
where a subscript colon ``;'' is used according to (\ref{eq:cov_T}).  Under the 
usual assumption that $\boldsymbol{\eta}(u,t)\to 0$ as $\vert u\vert 
\to\infty$, Stokes theorem then gives
\begin{equation}
	\int_{\mathbb{R}^2}\boldsymbol{\eta}_{,\beta}\cdot\boldsymbol{R}_{,\alpha}\,\pi^{\alpha\beta}\,\mathrm{vol}^2
	=-\int_{\mathbb{R}^2} 
	\boldsymbol{\eta}\cdot\left(\boldsymbol{R}_{,\alpha}\,{\pi^{\alpha\beta}}_{;\beta} 
	+ 
	\hat{\boldsymbol{n}}\,\pi^{\alpha\beta}\,b_{\alpha\beta}\right)\mathrm{vol}^2.
	\label{eq:Stokes_1st}
\end{equation}
For the second term on the right-hand side of (\ref{eq:deltaF_1}), consider the 
1-form
\begin{equation}
	\xi =
	\boldsymbol{\eta}_{,\alpha}\cdot\hat{\boldsymbol{n}}\,\psi^{\alpha\beta}\,i_{\vec{e}_\beta}\,\mathrm{vol}^2,
	\label{eq:xi}
\end{equation}
which, via \S \ref{app:int}, implies
\begin{equation}
	\mathrm{d}\xi = 
	\boldsymbol{\eta}_{,\alpha\beta}\cdot\hat{\boldsymbol{n}}\,\psi^{\alpha\beta}\,\mathrm{vol}^2
	+\boldsymbol{\eta}_{,\alpha}\cdot
	\left[
		\hat{\boldsymbol{n}}
		\left(
			{\psi^{\alpha\beta}}_{;\beta}
			- \psi^{\mu\nu}\,\Gamma^\alpha_{\mu\nu}
		\right)	- 
		\boldsymbol{R}_{,\gamma}\,{b^\gamma}_\beta\,\psi^{\alpha\beta}
	\right]\,\mathrm{vol}^2,
	\label{eq:dxi}
\end{equation}
and therefore, under the assumption that $\boldsymbol{\eta}_{,\alpha}(u,t)\to 
0$ as $\vert u\vert \to\infty$, applying Stokes theorem gives the result 
\begin{equation}
	\int_{\mathbb{R}^2} 
	\boldsymbol{\eta}_{,\alpha;\beta}\cdot\hat{\boldsymbol{n}}\,\psi^{\alpha\beta} 
	\mathrm{vol}^2 = \int_{\mathbb{R}^2} \boldsymbol{\eta}_{,\alpha}\cdot
	\left(
		\boldsymbol{R}_{,\gamma}\,{b^\gamma}_\beta\,\psi^{\alpha\beta}
		- \hat{\boldsymbol{n}}\,{\psi^{\alpha\beta}}_{;\beta}
		\right)\,\mathrm{vol}^2,
		\label{eq:Stokes_xi}
\end{equation}
where we have used the fact that $\boldsymbol{\eta}_{,\alpha;\beta}= 
\boldsymbol{\eta}_{,\alpha\beta} - 
\boldsymbol{\eta}_{,\gamma}\,\Gamma^\gamma_{\alpha\beta}$.  The final step is 
to apply Stokes theorem to the right-hand side of (\ref{eq:Stokes_xi}).  To 
this end, consider the 1-form
\begin{equation}
	\zeta = \boldsymbol{\eta}\cdot
	\left(
		\boldsymbol{R}_{,\gamma}\,{b^\gamma}_\beta\,\psi^{\alpha\beta}
		- \hat{\boldsymbol{n}}\,{\psi^{\alpha\beta}}_{;\beta}
		\right)\,i_{\vec{e}_\alpha}\,\mathrm{vol}^2,
		\label{eq:zeta}
\end{equation}
which, via \S \ref{app:int}, implies
\begin{equation}
	\mathrm{d}\zeta = \boldsymbol{\eta}_{,\alpha}\cdot
		\left(
			\boldsymbol{R}_{,\gamma}\,{b^\gamma}_\beta\,\psi^{\alpha\beta}
			- \hat{\boldsymbol{n}}\,{\psi^{\alpha\beta}}_{;\beta}
		\right)
		\,\mathrm{vol}^2
		+
		\boldsymbol{\eta}\cdot
		\left[
			\boldsymbol{R}_{,\alpha}
			\left(
			{b^\alpha}_\gamma\,\psi^{\gamma\beta}
			\right)_{;\beta}
			+
			\hat{\boldsymbol{n}}\,{b^\alpha}_\gamma\,\psi^{\gamma\beta}\,b_{\alpha\beta}
			+
			\boldsymbol{R}_{,\gamma}\,{b^\gamma}_\alpha\,{\psi^{\alpha\beta}}_{;\beta}
			-
			\hat{\boldsymbol{n}}\,{\psi^{\alpha\beta}}_{;\alpha\beta}
		\right]\,\mathrm{vol}^2.
		\label{eq:dzeta}
\end{equation}
After a final application of Stokes Theorem, the result can be combined with 
(\ref{eq:Stokes_1st}) and (\ref{eq:Stokes_xi}) to identify the functional 
derivative via
\begin{equation}
	\delta\mathcal{F} = \epsilon\int_{\mathbb{R}^2}\boldsymbol{\eta}\cdot\left( 
	\frac{\delta\mathcal{F}}{\delta\boldsymbol{R}} \right)\,\mathrm{vol}^2,
	\label{eq:functional}
\end{equation}
which gives
\begin{equation}
	\frac{\delta\mathcal{F}}{\delta\boldsymbol{R}} = \left[ 
	-{\pi^{\alpha\beta}}_{;\beta} - \left( 
	{b^\alpha}_\gamma\,\psi^{\gamma\beta} \right)_{;\beta} - 
	{b^\alpha}_\gamma\,{\psi^{\gamma\beta}}_{;\beta} 
	\right]\,\boldsymbol{R}_{,\alpha} + \left[ 
		-\pi^{\alpha\beta}\,b_{\alpha\beta} - \psi^{\alpha\beta}\left( 
	2H\,b_{\alpha\beta} - K\,g_{\alpha\beta} \right) + 
{\psi^{\alpha\beta}}_{;\alpha\beta} \right]\,\hat{\boldsymbol{n}}.
	\label{eq:detaF_deltaR_SI}
\end{equation}

\section{Dissipation}\label{sec:dis}
We may estimate the relative rates of energy dissipation that can be attributed 
to the tissue and the embedding fluid, respectively, during gastrulation.  The 
former arises not from relative movements of cells, but due to viscous shear of 
the enclosed cytosol as cells are deformed.  The latter arises due to viscous 
shear of the yolk.

We imagine the embryo as a sphere, parameterised in the usual spherical 
coordinates: radius $R$, azimuthal angle $\phi$, and polar angle $\theta$.  We 
consider a surface flow $\boldsymbol{v} = 
V\,\sin\theta\,\hat{\boldsymbol{\phi}}$.  In the bulk, assuming a no-slip 
boundary condition, geometric constraints impose a characteristic velocity 
gradient $V/R$.  An order of magnitude estimate of the rate of energy 
dissipation $\partial_t\,E_\mathrm{bulk}$ is therefore given by
\begin{equation}
	\partial_t\,E_\mathrm{bulk} \sim \eta_\mathrm{bulk}\,\left( \frac{V}{R} 
	\right)^2 \,\frac{4}{3}\,\pi\,R^3.
	\label{eq:E_dot_b}
\end{equation}
By contrast, on the surface, the characteristic velocity gradient is 
$2\,V/\pi\,R$.  Implying 
\begin{equation}
\partial_t\,E_\mathrm{surf} \sim \eta_\mathrm{surf}\,\left( \frac{2\,V}{\pi\,R} 
\right)^2 \,4\,\pi\,R^2\,\ell,
	\label{eq:E_dot_s}
\end{equation}
where $\ell$ is the thickness of the epithelium.  As a result
\begin{equation}
	\frac{\partial_t\,E_\mathrm{bulk}}{\partial_t\,E_\mathrm{surf}} \sim 
	\frac{\eta_\mathrm{bulk}}{\eta_\mathrm{surf}}\,\frac{R}{\ell}\,\frac{\pi^2}{12}.
	\label{eq:ratio_disip_gen}
\end{equation}
Assuming representative values $R = 200\,\mu \mathrm{m}$ and $\ell = 10\,\mu 
\mathrm{m}$ \cite{BA+89}, we may also use the fact that the meso-scale 
viscosity of of cellular cytosol is approximately 1 $\mathrm{Pa}\,\mathrm{s}$ 
\cite{Wirtz2009a}, as is the viscosity of embryonic yolk \cite{Wessel2015a}.  
For Drosophila, therefore, this gives
\begin{equation}
	\frac{\partial_t\,E_\mathrm{bulk}}{\partial_t\,E_\mathrm{surf}} \sim 
	2\,\pi^2,
	\label{eq:ratio_disip}
\end{equation}
implying that the dissipative contribution of the yolk is at least an order of 
magnitude greater than that of the epithelium.

\section{Active contractility}\label{sec:active}
An active contractile stress $\sigma = 
\sigma^{\alpha\beta}\,\boldsymbol{R}_{,\alpha}\otimes\boldsymbol{R}_{,\beta}$ 
acts in the tangent plane of $\mathcal{S}_t$.  The coefficients are of the form
\begin{equation}
	\sigma^{\alpha\beta} = \chi\left( \rho, \rho_\mathrm{b} 
	\right)\,\Delta\mu_{\mathrm{ATP}}\,g^{\alpha\beta}.
	%= \frac{\xi_1\,\rho_\mathrm{b}\,\chi\left( \rho 
	%\right)\,\Delta\mu_{\mathrm{ATP}}}{\left( 1+\xi_2\,\rho_{\mathrm{b}} 
%\right)}\,g^{\alpha\beta}.
	\label{eq:active_SI}
\end{equation}
Using the covariant analogue of the gradient operator 
$\nabla=\boldsymbol{R}_{,\mu}\,g^{\mu\nu}\,\partial_\nu$, the corresponding 
force (per unit area) is given by
\begin{equation}
	\begin{split}
		\nabla\cdot\sigma &= {\sigma^{\alpha\beta}}_{,\nu}\,g^{\mu\nu}\,\left( 
		\boldsymbol{R}_{,\mu}\cdot\boldsymbol{R}_{,\alpha} 
		\right)\,\boldsymbol{R}_{,\beta} + 
		\sigma^{\alpha\beta}\,g^{\mu\nu}\,\left(\boldsymbol{R}_{,\mu}\cdot\boldsymbol{R}_{,\alpha\nu} 
		\right)\,\boldsymbol{R}_{,\beta} + 
		\sigma^{\alpha\beta}\,g^{\mu\nu}\,\left(\boldsymbol{R}_{,\mu}\cdot\boldsymbol{R}_{,\alpha} 
		\right)\,\boldsymbol{R}_{,\beta\nu}\\
		&= {\sigma^{\alpha\beta}}_{,\alpha}\,\boldsymbol{R}_{,\beta} + 
		\sigma^{\alpha\beta}\,\Gamma^{\nu}_{\alpha\nu}\,\boldsymbol{R}_{,\beta} + 
		\sigma^{\alpha\beta}\,\left( 
		\Gamma^\gamma_{\alpha\beta}\,\boldsymbol{R}_{,\gamma} + 
		b_{\alpha\beta}\,\hat{\boldsymbol{n}} \right)\\
		&= {\sigma^{\alpha\beta}}_{;\alpha}\,\boldsymbol{R}_{,\beta} + 
		\sigma^{\alpha\beta}\,b_{\alpha\beta}\,\hat{\boldsymbol{n}},
		\label{eq:nabla_dot_sigma}
	\end{split}
\end{equation}
where the Gauss relation (\ref{eq:Gauss}) and the definition of the covariant 
derivative have been used.

Since the active stress relies on the density of ``bound'' or ``activated'' 
myosin, an auxiliary equation for the time dependence of $\rho_\mathrm{b}(u,t) 
= m_\mathrm{b} (u,t) / \sqrt{g} (u,t)$ is needed (here, $m_\mathrm{b}$ is the 
local mass of Myosin at time $t=0$).  We write the continuity equation for a 
conserved scalar field on a deformable manifold, including Langmuir-like source 
and sink terms:
\begin{equation}
	\partial_t \rho_b + \left( \rho_b \,v^{\alpha} \right)_{;\alpha} - 
	\rho_b\,v^{(n)}\,2H =
	%D\Delta\rho_b k_b\,\rho
	k_{\mathrm{on}} - k_{\mathrm{off}}\rho_b.
	%e^{ \mathsf{A}:\mathsf{\epsilon}},
	\label{eq:CoMy_SI}
\end{equation}
At this stage we do not specify the functional dependence of $k_\mathrm{on}$ 
and $k_\mathrm{off}$ which may be related to mechanical properties such as 
strains, strain-rates, stresses or forces (gradients of stress).  In the 
example calculation, both $k_\mathrm{on}$ and $k_\mathrm{off}$ are taken to be 
constant.  [Note that, unlike (\ref{eq:partial_rho_SI}), the convective term 
$v^\alpha\left(\rho_\mathrm{b}\right)_{;\alpha}$ cannot be ignored, even if the 
initial mass distribution (at $t=0$) was homoegeneous, due to the presence of 
source and sink terms].

\section{Quasi-1D example}\label{sec:Q1D}

\subsection{Setup}
Consider the following ``quasi-1D'' example.  At time $t=0$, the tissue is in a  
flat steady state ({\it i.e.}, $b_{\alpha\beta}(u,0) = 0$ for all $u$).  
Although flat, the tissue is active: there is a homogeneous steady state 
concentration of bound myosin $\rho_\mathrm{b}(u,0)$, which leads to 
homogeneous contractile stresses $\sigma$.  If there is no opposing force 
applied at the tissue boundary, then the contractile stresses are balanced by 
the restoring forces associated with the mechanical response of the cells.  

For convenience, we choose to apply a force at the boundary such that the 
(homogeneous) metric at steady state, $g_{\alpha\beta}(u,0)$, is the same as 
the rest metric $g^\dagger_{\alpha\beta}$.  We are then free to choose the 
coordinates $u$ such that $g_{\alpha\beta}(u,0) = g^\dagger_{\alpha\beta} = 
\delta_{\alpha\beta}$.  [Note that, formally, we say $\mathcal{S}_{t=0}$ is 
	isometric to $\mathbb{R}^2$.  This permits the decomposition 
	$\boldsymbol{R}_t=\Gamma_t\circ\gamma$, where 
	$\gamma:\mathbb{R}^2\to\mathbb{R}^3$ maps points $u$ to $\mathcal{S}_0$, 
	and $\Gamma_t:\mathbb{R}^3\to\mathbb{R}^3$ is a diffeomorphism between 
	positions of a give point $u$ at different times.  The coefficients 
	$\epsilon_{\alpha\beta}$ are then just given by 
	$\boldsymbol{R}_t^\ast\,\epsilon^{\mathbb{R}^3}\left(\vec{e}_\alpha,\vec{e}_{\beta} 
	\right) = g_{\alpha\beta}-\delta_{\alpha\beta}$, where 
	$\epsilon^{\mathbb{R}^3} = \left[ \left( \psi_0^\ast - \psi_{-t}^\ast 
	\right)\mathrm{I} \right]$ is the Green-Lagrange strain in $\mathbb{R}^3$, 
	and $\mathrm{I} = \delta_{ij}\,\mathrm{d}x^i\otimes\mathrm{d}x^j$ is the 
first fundamental form of three dimensional Euclidean space.].

Using a Monge-like parameterisation, we then consider the linear response of 
the tissue to small perturbations.  We define the three functions $R_t^i$ that 
comprise the embedding $\boldsymbol{R}_t:\mathbb{R}^2\to\mathbb{R}^3$, as:
\begin{equation}
	x^1 = R_t^1(u) = u^1 + \varepsilon\,\omega(u^1, t),\ x^2 = R_t^2(u) = u^2,\ 
	\mathrm{and} \ \mathrm{and}\ x^3 = R_t^3(u) = \varepsilon\,h(u^1,t),
	\label{eq:R_t_monge}
\end{equation}
where $\varepsilon\ll 1$ is a small dimensionless number.  For notational 
simplicty, let $u^1$ and $u^2$ be replaced by Euclidean $x$ and $y$, 
respectively.  We have
\begin{equation}
	\boldsymbol{R}(x,y,t) =
	\begin{pmatrix}
		x + \varepsilon\,\omega(x,t)\\
		y\\
		\varepsilon\,h(x,t)
	\end{pmatrix},\ \boldsymbol{R}_{,1} = \begin{pmatrix}
		1+\varepsilon\,\partial_x \omega\\
		0\\
		\varepsilon\,\partial_x h
	\end{pmatrix},\ \mathrm{and}\  \boldsymbol{R}_{,2} = \begin{pmatrix}
		0\\
		1\\
		0
	\end{pmatrix}.\label{eq:R_monge}
\end{equation}
To linear order,
\begin{equation}
	g_{\alpha\beta}\simeq
	\begin{pmatrix}
		1+2\,\varepsilon\,\partial_x\omega & 0\\
		0 & 1
	\end{pmatrix},\ g^{\alpha\beta}\simeq
	\begin{pmatrix}
		1-2\,\varepsilon\,\partial_x\omega & 0\\
		0 & 1
	\end{pmatrix},\ \mathrm{and}\ \hat{\boldsymbol{n}} \simeq \begin{pmatrix}
		-\varepsilon\,\partial_x h\\
		0\\
		1
	\end{pmatrix}.\label{eq:met_monge}
\end{equation}
Moreover, since
\begin{equation}
	\boldsymbol{R}_{,11} = \begin{pmatrix}
		\varepsilon\,\partial^2_x \omega\\
		0\\
		\varepsilon\,\partial^2_x h
	\end{pmatrix},\ \mathrm{and}\ \boldsymbol{R}_{12} = \boldsymbol{R}_{21} = 
	\boldsymbol{R}_{22} = 0,
	\label{eq:R_comma_monge}
\end{equation}
then
\begin{equation}
	b_{\alpha\beta} \simeq \begin{pmatrix} \varepsilon\,\partial^2_x h & 0\\
		0 & 0 \end{pmatrix}\ \Longrightarrow\ 
	H\simeq\frac{\varepsilon\,\partial_x^2 h}{2},\ \mathrm{and}\ K\simeq 0.
	\label{eq:b_monge}
\end{equation}

\subsection{Mechanical response}\label{sec:passive_monge}
Working to lowest order in $\varepsilon$, the apical strains are given by
\begin{equation}
	\epsilon_{\alpha\beta} \simeq \epsilon^{\alpha\beta}\simeq
	\begin{pmatrix}
		2\,\varepsilon\,\partial_x\omega & 0\\
		0 & 0
	\end{pmatrix},\ \mathrm{Tr}_g\left( \epsilon_{\alpha\beta} \right) =  
	\mathrm{Tr}_g\left( \epsilon^{\alpha\beta} \right)\simeq 
	2\,\varepsilon\,\partial_x\omega,\ \mathrm{and}\ 
	\bar{\epsilon}_{\alpha\beta} \simeq \bar{\epsilon}^{\alpha\beta}\simeq
	\begin{pmatrix}
		\varepsilon\,\partial_x\omega & 0\\
		0 & -\varepsilon\,\partial_x\omega\end{pmatrix},
	\label{eq:strain_monge_A}
\end{equation}
where $\bar{\epsilon}_{\alpha\beta} = \epsilon_{\alpha\beta} - 
g_{\alpha\beta}\,\mathrm{Tr}_g\left( \epsilon_{\alpha\beta} \right) / 2$ is the 
traceless-symmetric part of $\epsilon_{\alpha\beta}$.  The overall contribution 
of apical faces to the effective free energy is, to lowest order,
\begin{equation}
	C^{\alpha\beta\gamma\delta}\,\epsilon_{\alpha\beta}\,\epsilon_{\gamma\delta} 
	= \mu\,\bar{\epsilon}_{\alpha\beta}\,\bar{\epsilon}^{\alpha\beta} + 
	\lambda\,\left[\mathrm{Tr}_g\left( \epsilon_{\alpha\beta} \right)\right]^2= 
	4\varepsilon^2\left( \lambda_\mathrm{A} + \frac{\mu_\mathrm{A}}{2} 
	\right)\left( \partial_x\omega \right)^2+O\left( \varepsilon^3 \right).
	\label{eq:apical_elas}
\end{equation}
Due to the volume constraint (\ref{eq:V_constraint}), we have
\begin{equation}
	\ell = V\left( 1-2\,\varepsilon\,\partial_x\omega \right) + O\left( 
	\varepsilon^2 \right),
	\label{eq:V_approx}
\end{equation}
which assumes $\left\vert\ell\,\partial_x^2 h\right\vert / 2 = \varepsilon \ll 
1$.  Substituting into the expressions for the basal metric and inverse metric 
[Eqs.~(\ref{eq:g_basal}) and (\ref{eq:g_basal_inverse}), respectively] and 
keeping only the terms to linear order in $\varepsilon$, the resultant strains 
are
\begin{equation}
	\epsilon^\mathrm{B}_{\alpha\beta} \simeq \epsilon^{\alpha\beta}\simeq
	\begin{pmatrix}
		\varepsilon\,2\,\left(\partial_x\omega + V\,\partial^2_x h\right) & 0\\
		0 & 0
	\end{pmatrix},\ \mathrm{Tr}_g\left( \epsilon_{\alpha\beta}^\mathrm{B} 
	\right) =  \mathrm{Tr}_g\left( \epsilon^{\alpha\beta}_\mathrm{B} 
	\right)\simeq \varepsilon\,2\,\left(\partial_x\omega + V\,\partial^2_x 
	h\right),\ 
	%\mathrm{and}\ \bar{\epsilon}_{\alpha\beta} \simeq 
	%\bar{\epsilon}^{\alpha\beta}\simeq
	%\begin{pmatrix}
		%\varepsilon\left(\partial_x\omega - \frac{V\,\partial^2_x h}{2}\right) 
		%& 0\\
		%0 & -\varepsilon\left(\partial_x\omega - \frac{V\,\partial^2_x 
		%h}{2}\right)\end{pmatrix},
	\label{eq:strain_monge_B}
\end{equation}
and
\begin{equation}
	\bar{\epsilon}_{\alpha\beta}^\mathrm{B} \simeq 
	\bar{\epsilon}^{\alpha\beta}_\mathrm{B}\simeq
	\begin{pmatrix}
		\varepsilon\left(\partial_x\omega + \frac{V\,\partial^2_x h}{2}\right) 
		& 0\\
		0 & -\varepsilon\left(\partial_x\omega + \frac{V\,\partial^2_x 
		h}{2}\right)\end{pmatrix}.
	\label{eq:strain_monge_B_2}
\end{equation}
The overall contribution of basal elasticity to the free energy is, to lowest 
order, 
\begin{equation}
C^{\alpha\beta\gamma\delta}_\mathrm{B}\,\epsilon^\mathrm{B}_{\alpha\beta}\,\epsilon^\mathrm{B}_{\gamma\delta} 
= 
\mu_\mathrm{B}\,\bar{\epsilon}^\mathrm{B}_{\alpha\beta}\,\bar{\epsilon}_\mathrm{B}^{\alpha\beta} 
+ \lambda_\mathrm{B}\,\left[\mathrm{Tr}_g\left( 
\epsilon^\mathrm{B}_{\alpha\beta} \right)\right]^2= \varepsilon^2\left( 
\lambda_\mathrm{B} + \frac{\mu_\mathrm{B}}{2} \right)\,4\,\left( 
\partial_x\omega + V\,\partial_x^2 h\right)^2+O\left( \varepsilon^3 \right).
	\label{eq:basal_elas}
\end{equation}
We may then compute $\pi$ and $\psi$ according to (\ref{eq:pi}) and 
(\ref{eq:Lambda_varphi}), giving
\begin{equation}
	\pi^{11} = \frac{\kappa}{2}\left[ \left( \ell^\dagger \right)^2 - V^2 
	\right] +
	4\,\varepsilon\,\bigg\{\partial_x\omega\,\left( \lambda_\mathrm{A} + 
	\lambda_\mathrm{B} + \frac{\mu_\mathrm{A}}{2}  + \frac{\mu_\mathrm{B}}{2} + 
	\frac{\kappa}{4}\left[ 2\,V^2 -\left( \ell^\dagger \right)^2 \right]\right) 
	+ V\,\partial_x^2 h\,\left(  \lambda_\mathrm{B} + 
	\frac{\mu_\mathrm{B}}{2}\right)\bigg\} + O\left( \varepsilon^2 \right),
	\label{eq:pi_monge}
\end{equation}
and
\begin{equation}
	\psi^{11} = 8\,\varepsilon\,V\,\left[ \left(V\,\partial_x^2 h + 
	2\,\partial_x\omega\right)\left( \lambda_\mathrm{B} + 
\frac{\mu_\mathrm{B}}{2} \right) \right] + O\left( \varepsilon^2 \right),
	\label{eqpsi_monge}
\end{equation}
respectively, where $\pi^{22} = \pi^{12} = \pi^{21} = \psi^{22} = \psi^{12} = 
\psi^{21} = 0$.  Writing $\boldsymbol{f}_\mathrm{el} = -\delta \mathcal{F} / 
\delta \boldsymbol{R}$ and invoking (\ref{eq:detaF_deltaR_SI}), we see that the 
components of the resulting passive forces are
\begin{equation}
	f_\mathrm{el}^1 = 4\,\varepsilon\,\bigg\{\partial^2_x\omega\,\left( 
		\lambda_\mathrm{A} + \lambda_\mathrm{B} + \frac{\mu_\mathrm{A}}{2}  + 
		\frac{\mu_\mathrm{B}}{2} + \frac{\kappa}{4}\left[ 2\,V^2 -\left( 
		\ell^\dagger \right)^2 \right]\right) + V\,\partial_x^3 h\,\left(  
		\lambda_\mathrm{B} + \frac{\mu_\mathrm{B}}{2}\right)\bigg\} + O\left( 
		\varepsilon^2 \right),
	\label{eq:f_el^1_monge}
\end{equation}
and
\begin{equation}
	f_\mathrm{el}^{(n)} = \varepsilon\bigg\{\frac{\kappa}{2}\left[ \left( 
	\ell^\dagger \right)^2 - V^2 \right]\partial^2_x h - 8\,V\,\left[ 
	\left(V\,\partial_x^4 h + \partial^3_x\omega\right)\left( 
\lambda_\mathrm{B} + \frac{\mu_\mathrm{B}}{2} \right) \right]\bigg\} + O\left( 
\varepsilon^2 \right),
	\label{eq:f_el^n_monge}
\end{equation}
where $f_\mathrm{el}^2 = 0$.

\subsection{Active contractilty}\label{sec:active_monge}
Writing $\rho (x,t)=\rho^{(0)} + \varepsilon\,\rho^{(1)}(x,t) + O\left( 
\varepsilon^2 \right)$ and $\rho_\mathrm{b}(x,t)=\rho^{(0)}_\mathrm{b} + 
\varepsilon\,\rho^{(1)}_\mathrm{b}(x,t) + O\left( \varepsilon^2 \right)$, 
Eqn.~(\ref{eq:active_SI}) may be expanded as a power series in $\varepsilon$.  
Before doing so, we identify $\rho(x,0)$ with $\rho^{(0)}$,  {\it i.e.}, the 
steady state number density of cells per unit area is the same as that at time 
$t=0$.  From here, it is straightforward to show that $\rho^{(1)} = 
-\rho^{(0)}\,\partial_x\omega$ (see \S\ref{sec:rho}).  Similarly, from the 
continuity equation for $\rho_\mathrm{b}$ [Eq.~(\ref{eq:CoMy_SI})]
%
%\begin{equation}
	%\partial_t \rho_b + \rho_b \,v^{\alpha}_{;\alpha} - \rho_b\,v^{(n)}\,2H =
	%%D\Delta\rho_b k_b\,\rho
	%k_{\mathrm{on}} - k_{\mathrm{off}}\rho_b,
	%%e^{ \mathsf{A}:\mathsf{\epsilon}},
	%\label{eq:CoMy}
%\end{equation}
%
we see that $\rho_\mathrm{b}^{(0)} = k_{\mathrm{on}} / k_{\mathrm{off}} =: k$.  
Expanding (\ref{eq:active_SI}), we have
\begin{equation}
	\sigma^{\alpha\beta} = 
	\chi^{(0)}\,\Delta\mu_{\mathrm{ATP}}\,\delta^{\alpha\beta}
	+ \varepsilon\,\Delta\mu_{\mathrm{ATP}}\,\bigg[\chi^{(0)}\,\left( 
		g^{\alpha\beta} \right)^{(1)} + 
		\left.\frac{\partial\chi}{\partial\rho_\mathrm{b}}\right\vert_{\varepsilon=0}
		\,\rho_\mathrm{b}^{(1)}\,\delta^{\alpha\beta}
-\left.\frac{\partial\chi}{\partial\rho}\right\vert_{\varepsilon=0}\,\rho^{(0)}\,\partial_x\omega\,\delta^{\alpha\beta} 
\bigg] + O\left( \varepsilon^2 \right),
	\label{eq:sigma_monge}
\end{equation}
where $\chi^{(0)}=\chi^{(0)}\left( \rho^{(0)},k \right)$.  Writing 
$\boldsymbol{f}_\mathrm{ac} = \nabla\cdot\sigma$ and invoking 
(\ref{eq:nabla_dot_sigma}), the components of the resulting active forces are
\begin{equation}
	f^{1}_\mathrm{ac} =\varepsilon\,\Delta\mu_{\mathrm{ATP}} \left[ 
		\partial_x\rho_\mathrm{b}^{\left( 0 
		\right)}\left.\frac{\partial\chi}{\partial\rho_{\mathrm{b}}}\right\vert_{\varepsilon=0}
	- \partial_x^2\omega\,\left( \chi^{(0)} +  \rho^{(0)}\,
	\left.\frac{\partial\chi}{\partial\rho}\right\vert_{\varepsilon=0}\right)\right]+ 
	O\left( \varepsilon^2 \right),\label{eq:f^1_ac}
\end{equation}
and
\begin{equation}
	f^{(n)}_\mathrm{ac} =  
	\varepsilon\,\Delta\mu_{\mathrm{ATP}}\,\chi^{(0)}\,\partial_x^2 h+ O\left( 
	\varepsilon^2 \right),\label{eq:f^n_ac}
\end{equation}
where $f^{2}_\mathrm{ac} = 0$.

\subsection{Embedding fluid}
The epithelium is impermeable, and we therefore assume a ``no-slip'' condition 
between $\mathcal{S}_t$ and
the embedding fluid.  The movement of $\mathcal{S}_t$, specified by the 
velocities $\boldsymbol{v}(u,t)=\partial\boldsymbol{R}(u,t)/\partial t$, is 
therefore related to the forces $\boldsymbol{f} =  \boldsymbol{f}_\mathrm{el} + 
\boldsymbol{f}_\mathrm{ac}$ (\S\ref{sec:passive_monge} and 
\S\ref{sec:active_monge}) via convolution with the Oseen tensor 
\cite{HappelBrenner}:
\begin{equation}
	\boldsymbol{v}\left( \boldsymbol{R}\left( u,t \right) \right) = 
	\int_{\mathbb{R}^2}
\mathrm{d}u^\prime\,\mathsf{O}\left(  \boldsymbol{R}\left( u^\prime,t \right) - 
\boldsymbol{R}\left( u,t \right)\right)\cdot\boldsymbol{f}\left( 
\boldsymbol{R}\left( u^\prime,t \right) \right),
	\label{eq:oseen_SI}
\end{equation}
where the components of $\mathsf{O}$ in the local basis 
$\boldsymbol{e}_i=\left\{\boldsymbol{R}_{,1},\boldsymbol{R}_{,2},\hat{\boldsymbol{n}}\right\}$, 
are given by $\mathsf{O}^{ij}= 
\Lambda^i_p\,\Lambda_p^j\,\mathsf{O}^{pq}_{\mathbb{E}^3}$.  Here, 
$\mathsf{O}^{ij}_{\mathbb{E}^3}$ are the usual components of the Oseen tensor 
\begin{equation}
	\mathsf{O}_{\mathbb{E}^3}^{ij}(\boldsymbol{x}) = 
	\frac{1}{8\,\pi\,\eta\,\left\vert\boldsymbol{x}\right\vert}\left( 
	\delta^{ij}+ \frac{x^i x^j}{\left\vert\boldsymbol{x}\right\vert^2} \right),
	\label{eq:oseen_ij}
\end{equation}
where $\boldsymbol{x} = x^i\,\hat{\boldsymbol{e}}^{\mathbb{E}^3}_i$, such that 
$\hat{\boldsymbol{e}}^{\mathbb{E}^3}_i = 
\{\hat{\boldsymbol{x}},\hat{\boldsymbol{y}},\hat{\boldsymbol{z}}\}$
is the usual Euclidean basis.  The matrix $\Lambda$ is prescribed by the 
relation $\boldsymbol{e}_i = 
\Lambda_i^j\,\hat{\boldsymbol{e}}^{\mathbb{E}^3}_j$, and can be computed given 
an explicit embedding.  Using (\ref{eq:R_t_monge}) results in a power series 
expansion of $\mathsf{O}$ in terms $\varepsilon$.  However, since 
$\boldsymbol{f}$ has no $O(\mathrm{const.})$ term [ensuring that 
$\boldsymbol{v}=0$ when $\varepsilon=0$, {\it i.e.}, at steady state] then the 
$O(\varepsilon)$ contribution to $\boldsymbol{v}$ is just
\begin{equation}
	\boldsymbol{v}^{(1)}\left( \boldsymbol{R}\left( u,t \right) \right) = 
	\int_{\mathbb{R}^2}
	\mathrm{d}u^\prime\,\mathsf{O}^{(0)}\left(  \boldsymbol{R}\left( u^\prime,t 
	\right) - \boldsymbol{R}\left( u,t 
	\right)\right)\cdot\boldsymbol{f}^{(1)}\left( \boldsymbol{R}\left( 
	u^\prime,t \right) \right).
	\label{eq:oseen_monge}
\end{equation}
The lack of a zeroth order term also implies
\begin{equation}
	g^{\alpha\beta}\,\boldsymbol{R}_{,\beta}\cdot\boldsymbol{v} = 
	\boldsymbol{v}\cdot\hat{\boldsymbol{e}}^{\mathbb{E}^3}_\alpha + O\left( 
	\varepsilon^2 \right)\ \mathrm{and}\ 
	\hat{\boldsymbol{n}}\cdot\boldsymbol{v} = 
	\boldsymbol{v}\cdot\hat{\boldsymbol{e}}^{\mathbb{E}^3}_3+ O\left( 
	\varepsilon^2 \right).
	\label{eq:correspondence}
\end{equation}
That is, up to $O(\varepsilon^2)$, the Euclidean components of the velocity 
field are the same as those expressed in the basis local to $\mathcal{S}_t$.
As a result, we may write the following component-wise expression:
\begin{equation}
	\left[v^{(1)}\right]^i\left( x,y;t\right) = \int_{\mathbb{R}^2}
	\mathrm{d}u^\prime\,\left[\mathsf{O}^{(0)}\right]^i_j \left( x^\prime - 
	x,y^\prime-y \right)\,\left[{f}^{(1)}\right]^j\left( x^\prime, y^\prime;t 
	\right).
	\label{eq:oseen_monge_comp}
\end{equation}
where there is an implict sum over $j$, and the components of 
$\mathsf{O}^{(0)}$ are
\begin{equation}
	\left[\mathsf{O}^{(0)}\right]^i_j\left( x,y \right) = 
	\frac{1}{8\,\pi\,\eta\,\left( x^2+y^2 \right)^{1/2}}\left[
\begin{pmatrix}
		1& 0 &0\\
		0 & 1&0\\
		0&0&1
	\end{pmatrix}
	+ \frac{1}{\left( x^2+y^2 \right)}
\begin{pmatrix}
		xx& xy &0\\
		yx & yy&0\\
		0&0&0
	\end{pmatrix}
\right].
	\label{eq:oseen_ij^0}
\end{equation}
Using the notation 
\begin{equation}
	\mathscr{F}_{\boldsymbol{q}}
	\left\{\varphi \right\}=\int_{-\infty}^\infty\mathrm{d}x 
	\int_{-\infty}^\infty
	\mathrm{d}y\,\varphi\,e^{-i\,\boldsymbol{q}\cdot\boldsymbol{r}},
	\label{eq:FT}
\end{equation}
where $\boldsymbol{r} = (x,y)^\mathsf{T}$ and $\boldsymbol{q} = 
(q_x,q_y)^\mathsf{T}$, we may take the two-dimensional Fourier transform of 
(\ref{eq:oseen_monge_comp}).  The result is that
\begin{equation}
	\left[\mathscr{F}_{\boldsymbol{q}}
	\left\{ v^{(1)}\right\}\right]^i=\left[\mathscr{F}_{\boldsymbol{q}}
	\left\{ \mathsf{O}^{(0)} 
\right\}\right]^i_j\,\left[\mathscr{F}_{\boldsymbol{q}}
	\left\{ f^{(1)}\right\}\right]^j.
	\label{eq:FT_v}
\end{equation}
To compute the Fourier transform of the (zeroth order) Oseen tensor, we may 
exploit the fact that $\mathsf{O}^{(0)}_{ij} =\mathsf{O}_{\alpha\beta}^{(0)} 
\oplus\mathsf{O}_{33}^{(0)}$ and hence $\mathscr{F}_{\boldsymbol{q}}\left\{ 
	\mathsf{O}^{(0)} \right\} = \mathscr{F}_{\boldsymbol{q}}\left\{ 
		\mathsf{O}_{\alpha\beta}^{(0)} \right\} 
		\oplus\mathscr{F}_{\boldsymbol{q}}\left\{ \mathsf{O}_{33}^{(0)} 
	\right\}$, where 
\begin{equation}
	\mathsf{O}^{(0)}_{\alpha\beta}\left( x,y \right) = 
	\frac{1}{8\,\pi\,\eta\,\left\vert \boldsymbol{r} \right\vert}\left(
		\delta_{\alpha\beta} + \frac{r_\alpha\,r_\beta}{\left\vert 
			\boldsymbol{r} \right\vert^2}
\right),
	\label{eq:oseen_alphabeta^0}
\end{equation}
and $\mathsf{O}^{(0)}_{33}= 1/8\,\pi\,\eta\,\vert \boldsymbol{r}\vert$.  To 
compute the necessary Fourier transforms, we use the following three facts 
\cite{IntegralTransforms}:
\begin{enumerate}
	\item The $n$-dimensional Fourier transform of a homogeneous function of 
		degree $\lambda$ is a homogeneous function of degree $-\lambda-n$.
	\item The Fourier transform of a ``radial'' function ({\it i.e.}, depending 
		on distance rather than absolute position) is also a radial function.
	\item That $\mathscr{F}_{\boldsymbol{q}}  \left\{ \mathrm{Tr}\mathsf{O} 
	\right\} =\mathrm{Tr}\mathscr{F}_{\boldsymbol{q}} \left\{\mathsf{O} 
\right\}$.
\end{enumerate}
Consider first $\mathscr{F}_{\boldsymbol{q}}\left\{ 
	\mathsf{O}_{\alpha\beta}^{(0)} \right\}$, which, via linearity, is just the 
	sum of $\mathscr{F}_{\boldsymbol{q}}\left\{ 
		\frac{\delta_{\alpha\beta}}{8\,\pi\,\eta\,\vert\boldsymbol{r}\vert} 
	\right\}$ and $\mathscr{F}_{\boldsymbol{q}}\left\{ 
		\frac{r_\alpha\,r_\beta}{8\,\pi\,\eta\,\vert\boldsymbol{r}\vert^3} 
	\right\}$.  Using 1. and 2. above, we have,
\begin{equation}
	\mathscr{F}_{\boldsymbol{q}}\left\{ 
		\frac{\delta_{\alpha\beta}}{8\,\pi\,\eta\,\vert\boldsymbol{r}\vert} 
	\right\} = 
	\frac{\delta_{\alpha\beta}}{8\,\pi\,\eta}\int_{\mathbb{R}^2}\mathrm{d}\boldsymbol{r}\,e^{-i\,\boldsymbol{q}\cdot\boldsymbol{r}}\,\frac{1}{\vert\boldsymbol{r}\vert} 
	= 
	\frac{c_1\,\delta_{\alpha\beta}}{8\,\pi\,\eta\,\vert\boldsymbol{q}\vert},
	\label{eq:FT_1}
\end{equation}
and 
\begin{equation}
	\mathscr{F}_{\boldsymbol{q}}\left\{ 
		\frac{r_\alpha\,r_\beta}{8\,\pi\,\eta\,\vert\boldsymbol{r}\vert^3} 
	\right\} = 
	-\frac{\partial_{q_x}\,\partial_{q_y}}{8\,\pi\,\eta}\int_{\mathbb{R}^2}\mathrm{d}\boldsymbol{r}\,e^{-i\,\boldsymbol{q}\cdot\boldsymbol{r}}\,\frac{1}{\vert\boldsymbol{r}\vert^3} 
	= 
	-\frac{c_2\,\partial_{q_x}\,\partial_{q_y}\vert\boldsymbol{q}\vert}{8\,\pi\,\eta}
	= \frac{c_2}{8\,\pi\,\eta}\left( \delta_{\alpha\beta} - 
	\frac{q_\alpha\,q_\beta}{\vert\boldsymbol{q}\vert^2} \right).
	\label{eq:FT_2}
\end{equation}
The constants $c_1$ and $c_2$ can be fixed by invoking 3., above, which implies 
that $c_1=c_2=2\,\pi$, and hence
\begin{equation}
	\mathscr{F}_{\boldsymbol{q}}\left\{ \mathsf{O}_{\alpha\beta}^{(0)} \right\} 
	= \frac{1}{4\,\eta\,\vert\boldsymbol{q}\vert}\left( 2\,\delta_{\alpha\beta} 
	- \frac{q_\alpha\,q_\beta}{\vert\boldsymbol{q}\vert^2}\right).
	\label{eq:FT_1+2}
\end{equation}
Similarly, it is straightforward to show that 
$\mathscr{F}_{\boldsymbol{q}}\left\{ \mathsf{O}_{33}^{(0)} \right\} = 
1/4\,\eta\,\vert\boldsymbol{q}\vert$, and therefore
\begin{equation}
	\mathscr{F}_{\boldsymbol{q}}\left\{ \mathsf{O}_{ij}^{(0)} \right\} = 
	\frac{1}{4\,\eta\,\vert\boldsymbol{q}\vert}\left[
\begin{pmatrix}
		2& 0 &0\\
		0 & 2&0\\
		0&0&1
	\end{pmatrix}
	+ \frac{1}{\vert\boldsymbol{q}\vert^2}
\begin{pmatrix}
		q_x\,q_x& q_x\,q_y &0\\
		q_y\,q_x & q_y\,q_y&0\\
		0&0&0
	\end{pmatrix}
\right].
	\label{eq:FT_final}
\end{equation}
Turning our attention to $\mathscr{F}_{\boldsymbol{q}}
\left\{ \boldsymbol{f}^{(1)} \right\}$, we note that $\boldsymbol{f}^{(1)}$ is 
not a function of $y$, and therefore $\mathscr{F}_{\boldsymbol{q}}
\left\{ \boldsymbol{f}^{(1)} \right\} = \mathscr{F}_{q_x}
\left\{ \boldsymbol{f}^{(1)} \right\}\,\delta\left( -q_y / 2\pi \right)$.  
Dropping the subscript $x$ for convenience, and using the fact that 
$\left[\boldsymbol{f}^{(1)}\right]^2=0$, it can be shown that
\begin{equation}
	\mathscr{F}_{q}
	\left\{ 
		\left[\boldsymbol{v}^{(1)}\right]^i\right\}=\frac{1}{4\,\eta\,q}\mathscr{F}_{q}
			\left\{ \left[\boldsymbol{f}^{(1)}\right]^i\right\}.
	\label{eq:FT_v_final}
\end{equation}

\subsection{Linear stability}
For the general case of arbitrary deformations, there are five dynamical 
variables: two each for $g_{\alpha\beta}$ and $b_{\alpha\beta}$ (due to the 
Gauss-Codazzi relations) plus the scalar density $\rho_\mathrm{b}$.  However, 
when restricted to quasi-1D deformations, the number of degrees-of-freedom is 
reduced to three.  In our case, natural choices are the local surface area 
measure $\sqrt{g}$, the mean curvature $H$, and $\rho_\mathrm{b}$.  The 
corresponding dynamical equations are given by (\ref{eq:d_area_dt_result}), 
(\ref{eq:d_H_dt_result}) and (\ref{eq:CoMy_SI}).  Here, since $\boldsymbol{v} = 
\varepsilon\,\boldsymbol{v}^{(1)}(x,t) + O(\varepsilon^2)$ and 
$\hat{\boldsymbol{e}}_2^{\mathbb{E}^3}\cdot\boldsymbol{v}^{(1)}= 
\left[\boldsymbol{v}^{(1)}\right]^2 = 0$, at $O(\varepsilon)$ we have
\begin{equation}
	\partial_t \left( \sqrt{g} \right)^{(1)} = \partial_x \left[ 
	\boldsymbol{v}^{(1)} \right]^1\ \Longrightarrow\  -i\,q\,\partial_t 
	\mathscr{F}_q\left\{\omega\right\} = 
	-\frac{i}{4\,\eta}\mathscr{F}_q\left\{\left[ \boldsymbol{f}^{(1)} \right]^1 
\right\},
	\label{eq:lin_rootg_FT}
\end{equation}
\begin{equation}
	\partial_t H^{(1)} = \frac{1}{2}\partial^2_x \left[ \boldsymbol{v}^{(1)} 
	\right]^3\ \Longrightarrow\  -\frac{q^2}{2}\partial_t 
	\mathscr{F}_q\left\{h\right\} = 
	-\frac{q}{8\,\eta}\mathscr{F}_q\left\{\left[ \boldsymbol{f}^{(1)} \right]^3 
\right\},
	\label{eq:lin_H_FT}
\end{equation}
and
\begin{equation}
	\partial_t \rho_\mathrm{b}^{(1)} = -k\,\partial_x \left[ 
	\boldsymbol{v}^{(1)} \right]^1 - k_\mathrm{off}\,\rho_\mathrm{b}^{(1)}\ 
	\Longrightarrow\  \partial_t 
	\mathscr{F}_q\left\{\rho_\mathrm{b}^{(1)}\right\} = 
	\frac{i\,k}{4\,\eta}\mathscr{F}_q\left\{\left[ \boldsymbol{f}^{(1)} 
	\right]^1 \right\} - 
	k_\mathrm{off}\,\mathscr{F}_q\left\{\rho_\mathrm{b}^{(1)} \right\},
	\label{eq:lin_rhob_FT}
\end{equation}
where, introducing the shorthand $A:=\mu_A + \lambda_A$, $B:=\mu_B + 
\lambda_B$, $\alpha := 
\partial\chi/\partial\rho_\mathrm{b}\vert_{\varepsilon=0}$ and $\beta := 
\chi^{(0)} + \rho^{(0)}\,\partial\chi/\partial\rho\vert_{\varepsilon=0}$,
\begin{equation}
	\mathscr{F}_q\left\{\left[ \boldsymbol{f}^{(1)} \right]^1 \right\} = \left( 
	\beta\,\Delta\mu_\mathrm{ATP} - 4\left\{ A+B+\frac{\kappa}{4}\left[ 2\,V^2 
	- \left( \ell^\dagger \right)^2 \right] 
\right\}\right)\,q^2\,\mathscr{F}_q\left\{ \omega\right\} + 
i\,4\,q^3\,V\,B\,\mathscr{F}_q\left\{ h\right\} - 
i\,q\,\Delta\mu_\mathrm{ATP}\,\alpha\,\mathscr{F}_q\left\{\rho_\mathrm{b}^{(1)}\right\},
	\label{eq:FT_f^1}
\end{equation}
and
\begin{equation}
	\mathscr{F}_q\left\{\left[ \boldsymbol{f}^{(1)} \right]^3 \right\} = 
	-\left\{ \frac{\kappa}{2}\left[ \left( \ell^\dagger \right)^2 - V^2\right] 
	+ 8\,B\,V^2\,q^2 + 
	\Delta\mu_\mathrm{ATP}\,\chi^{(0)}\right\}\,q^2\,\mathscr{F}_q\left\{ 
	h\right\} -  8\,i\,q^3\,V\,B\,\mathscr{F}_q\left\{ \omega\right\}.
	\label{eq:FT_f^3}
\end{equation}
Eqs.~(\ref{eq:lin_rootg_FT}), (\ref{eq:lin_H_FT}) and (\ref{eq:lin_rhob_FT}) 
can be written in terms of the following dimensionless variables: $\Lambda:=A^2 
/ \eta^2\,\Delta\mu_{\mathrm{ATP}}$, $\Phi:=k\,A/\eta^2$, 
$\overline{\beta}:=\beta\,\Delta\mu_\mathrm{ATP} / A$, 
$\overline{\chi}:=\chi^{(0)}\,\Delta\mu_\mathrm{ATP} / A$, $\overline{q} := 
\ell^\dagger\,q$, $\overline{t} := 
\eta\,\Delta\mu_\mathrm{ATP}\,t/A\,\ell^\dagger$, $\gamma:=B/A$, 
$\overline{\kappa}:=\kappa\,\left(\ell^\dagger\right)^2 / A$, $\delta := 
V/\ell^\dagger$, $\overline{\mathscr{F}_q\left\{w\right\}} := 
\mathscr{F}_q\left\{w\right\}/\left( \ell^\dagger \right)^2$, 
$\overline{\mathscr{F}_q\left\{h\right\}} := 
\mathscr{F}_q\left\{h\right\}/\left( \ell^\dagger \right)^2$, and 
$\overline{\mathscr{F}_q\left\{\rho_\mathrm{b}\right\}} := 
\mathscr{F}_q\left\{\rho_\mathrm{b}\right\}/\beta\,\ell^\dagger$.  Writing 
$\boldsymbol{x} = \left( 
\overline{\mathscr{F}_q\left\{w\right\}},\overline{\mathscr{F}_q\left\{h\right\}},\overline{\mathscr{F}_q\left\{\rho_\mathrm{b}\right\}}\right)^\mathsf{T}$, 
the resulting matrix equation is given by 
$\partial_{\overline{t}}\,\boldsymbol{x}= \mathsf{M}\cdot\boldsymbol{x}$, where
%
%\begin{equation}
	%\partial_{\overline{t}}\,\boldsymbol{x}= \mathsf{M}\cdot\boldsymbol{x}, 	
	%\label{eq:lin_stab_matrix}
%\end{equation}
%
%
%\begin{equation}
	%\partial_{\overline{t}}\begin{pmatrix}
		%\overline{\mathscr{F}_q\left\{w\right\}}\\
		%\overline{\mathscr{F}_q\left\{h\right\}}\\
		%\overline{\mathscr{F}_q\left\{\rho_\mathrm{b}\right\}}
	%\end{pmatrix} = \mathsf{M}\cdot\begin{pmatrix}
		%\overline{\mathscr{F}_q\left\{w\right\}}\\
		%\overline{\mathscr{F}_q\left\{h\right\}}\\
		%\overline{\mathscr{F}_q\left\{\rho_\mathrm{b}\right\}}
	%\end{pmatrix}, 	\label{eq:lin_stab_matrix}
%\end{equation}
%
%where
%
\begin{equation}
	\mathsf{M} = \begin{pmatrix} \Lambda\,\overline{q}\,\left\{ 
			\overline{\beta} - 4\left[ 1+\gamma + \overline{\kappa}\left( 
			2\,\delta^2 -1 \right)/4 \right]\right\}/4 & + 
			i\,\Lambda\,\delta\,\gamma\,\overline{q}^2 & 
			-i\,\Lambda\,\alpha\,\overline{\chi}/4 \\
		-2\,i\,\Lambda\,\delta\,\gamma\,\overline{q}^2 
		&-\Lambda\,\overline{q}\,\left[ \overline{\beta} + 
		8\,\gamma\,\delta^2\,\overline{q}^2 + 
	\overline{\kappa}\,\left(1-\delta^2 \right)/2 \right]/4& 0\\
	i\,\Phi\,\overline{q}^2\,\left\{ \overline{\beta} - 4\left[ 1+\gamma + 
	\overline{\kappa}\left( 2\,\delta^2 -1 \right)/4 
\right]\right\}/\overline{\beta}& 
-\Phi\,\delta\,\gamma\,\overline{q}^3/\overline{\beta}& 
\Phi\,\alpha\,\overline{q} - \overline{k}_\mathrm{off}
	\end{pmatrix}.
\label{eq:M}
\end{equation}
Focussing only on the variables $\delta$, $\overline{\beta}$, and 
$\overline{\kappa}$, whose relative values can lead to sign changes of the 
coefficients of $\mathsf{M}$, we set all other variables equal to one ({\it 
i.e.}, 
$\Lambda=\Phi=\gamma=\overline{\chi}=\alpha=\overline{k}_\mathrm{off}=1$).  We 
may then solve the corresponding eigenvalue equation 
$\mathsf{M}\cdot\hat{\boldsymbol{E}}^{(i)}= 
\hat{\boldsymbol{E}}^{(i)}\,\lambda_i$.  Writing 
$\boldsymbol{x}(q,\overline{t}) = 
\sum_i\,\xi_i(q,\overline{t})\,\hat{\boldsymbol{E}}^{(i)}(q)$, and substituting 
into the aforementioned matrix equation, implies 
$\xi_i\left(q,\overline{t}\right) = e^{\lambda_i(q)\,\overline{t}}$, hence 
solving for $\boldsymbol{x}$.  To linear order in $\overline{q}$, the 
eigenvalues $\lambda_i$ ($i=1,2,3$) are real and given by:
\begin{equation}
	\lambda_\alpha = -\frac{\overline{q}}{16}\left[ 16 + \left( 3\,\delta^2 - 1 
	\right)\,\overline{\kappa} + (-1)^\alpha\,\left\vert 16 - 
4\,\overline{\beta} - 3\,\overline{\kappa} + 
5\,\delta^2\,\overline{\kappa}\right\vert \right], 
	\label{eq:lambda_1_2}
\end{equation}
for $\alpha=1,2$, and $\lambda_3 = \overline{q}-1$. 
%
%\begin{equation}
	%\lambda_3 = \overline{q}-1.
	%\label{eq:lambda_3}
%\end{equation}
%
The sign of (\ref{eq:lambda_1_2}), and hence the stability of the corresponding 
perturbation, changes according to two criteria
\begin{equation}
	\delta^2\,\overline{\kappa} - 2\,\overline{\beta} + \overline{\kappa} = 0, 
	\ \ \mathrm{and}\ \ 2\,\delta^2\,\overline{\kappa} + 8 - \overline{\beta} 
	-\overline{\kappa} = 0, \label{eq:stab_boundaries}
\end{equation}
which correspond to the red-solid and blue-dashed lines of Fig.~4{\bf a} of the 
main manuscript.  We consider three cases, each of which is characterised by a 
different instability.

In all cases, a single eigenvalue, $\lambda_1$, has the largest real part, 
irrespective of wave-number $\overline{q}$.  For a $\overline{q}$ corresponding 
to positive $\mathrm{Re}\left[\lambda_1\right]$, the resulting instability is 
characterised by the coefficients of the eigenvector $\hat{\boldsymbol{E}}^1$, 
both real and imaginary parts.  The relative growth rates of cosinusoidal 
perturbations $h(x,t)$, $\omega(x,t)$ and $\rho^{(1)}_\mathrm{b}(x,t)$ are 
given by the coefficients $\mathrm{Re}\left[E^1_2\right]$, 
$\mathrm{Re}\left[E^1_1\right]$, and $\mathrm{Re}\left[E^1_3\right]$, 
respectively.  Similarly, the relative growth rates of sinusoidal perturbations 
$h(x,t)$, $\omega(x,t)$ and $\rho^{(1)}_\mathrm{b}(x,t)$ are given by the 
coefficients $\mathrm{Im}\left[E^1_2\right]$, $\mathrm{Im}\left[E^1_1\right]$, 
and $\mathrm{Im}\left[E^1_3\right]$, respectively.

By analysing the fastest growing eigenvector in each of the cases of interest, 
the Monge parameterisation (\ref{eq:R_t_monge}) may be used to visualise the 
corresponding deformations of cell junctions.  Moreover, by invoking the 
cell-volume constraint (linearised, for consistency) to obtain cell thickness, 
we may construct a faithful representation of the tissue just after the onset 
of the instability (see Figs. \ref{fig:inv}, \ref{fig:squa}, and 
\ref{fig:buck}).

\begin{figure*}[h!]
\centering
\includegraphics[width=1.0\textwidth]{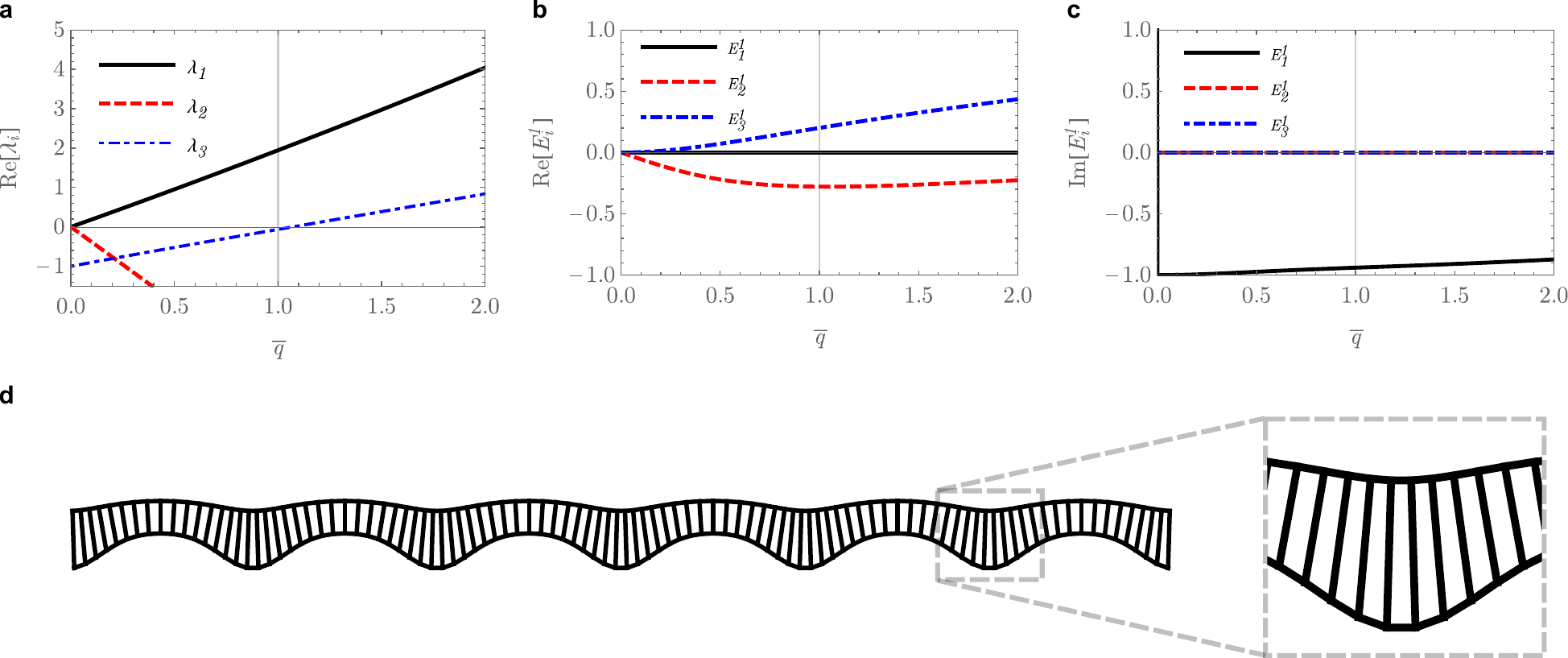}
%/home/rgm/Dropbox/Membranes/Epithelia/Elastomer/Figures/inv_lin_stab_sign.pdf
\caption
{
	(Color online).  Invagination: $\delta=\sqrt{2}$, $\overline{\beta}=15$, 
	$\overline{\kappa}=1/2$.  At linear order, the 
	$\mathrm{Re}\left[\lambda_1\right]$ instability is unbounded [panel {\bf 
	a})].  However, we expect non-liearities to provide an effective cutoff, 
	and therefore focus on characteristic behaviour at $\overline{q}=1$.  
	Cosinusoidal variations in $\rho_\mathrm{b}(x,t)$ panel {\bf b}) [blue 
	dot-dashed line] drive areas of apical contraction and expansion.  Since 
	$\delta >1$, the steady-state corresponds to lateral faces that are already 
	stretched.  As a result, contraction of the apical faces corresponds to 
	expansion of the basal faces (in addition to some further stretching of the 
	lateral faces), due to the conservation of cell volume.  At the cellular 
	scale this gives rise to regluar-prism to truncated-pyramid transitions 
	[panel {\bf d}) inset].  At a tissue level, this is seen as an invaginating 
	phase [panel {\bf d})] whose apical shape is a combination of both 
	sinusoidal variations in $\omega(x,t)$ [panel {\bf c}) black solid line] 
	and cosinusiodal variations in $h(x,t)$ [panel {\bf b}) red dashed line] 
	(each phase-shifted by a factor of $\pi$).
}
\label{fig:inv}
\end{figure*}
\begin{figure*}[!h]
\centering
\includegraphics[width=1.0\textwidth]{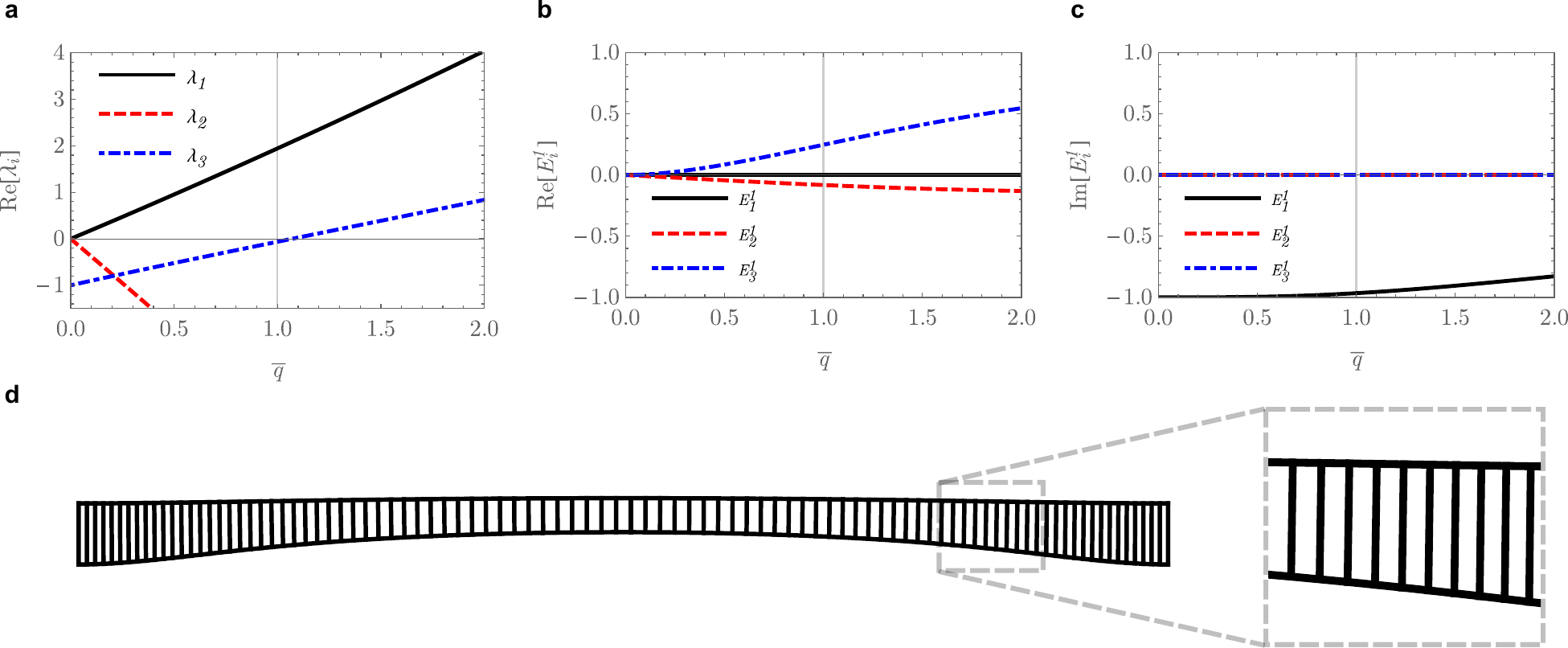}
%/home/rgm/Dropbox/Membranes/Epithelia/Elastomer/Figures/sqau_lin_stab_sign.pdf
\caption
{
	(Color online).  Sqaumous-to-columnar: $\delta=1/4$, $\overline{\beta}=15$, 
	$\overline{\kappa}=1/2$.  At linear order, the 
	$\mathrm{Re}\left[\lambda_1\right]$ instability is unbounded [panel {\bf 
	a})].  However, we expect non-liearities to provide an effective cutoff, 
	and therefore focus on characteristic behaviour at $\overline{q}=1$ 
	(although, for comparison with Fig.~\ref{fig:inv} the system has been 
	scaled so that cells occupy the same volume, even though $\delta$ is 
	different in the two cases).  Cosinusoidal variations in 
	$\rho_\mathrm{b}(x,t)$ [panel {\bf b}) blue dot-dashed line] drive areas of 
	apical contraction and expansion.  Since $\delta <1$, the steady-state 
	corresponds to lateral faces that are compressed.  As a result, contraction 
	of the apical faces corresponds to elongation of the lateral faces (in 
	addition to an almost imperceptible expansion of basal faces), due to the 
	conservation of cell volume.  At the cellular scale, this gives rise to 
	squamous to columnar transitions [panel {\bf d}) inset].  At a tissue 
	level, the apical shape is essentially flat [panel {\bf b}) red dashed 
	line], but with a thickness inversely proportional to $\pi$-phase-shifted 
	sinusoidal variations in $\omega(x,t)$ [panel {\bf c}) black solid line].
}
\label{fig:squa}
\end{figure*}
\begin{figure*}[t]
\centering
\includegraphics[width=1.0\textwidth]{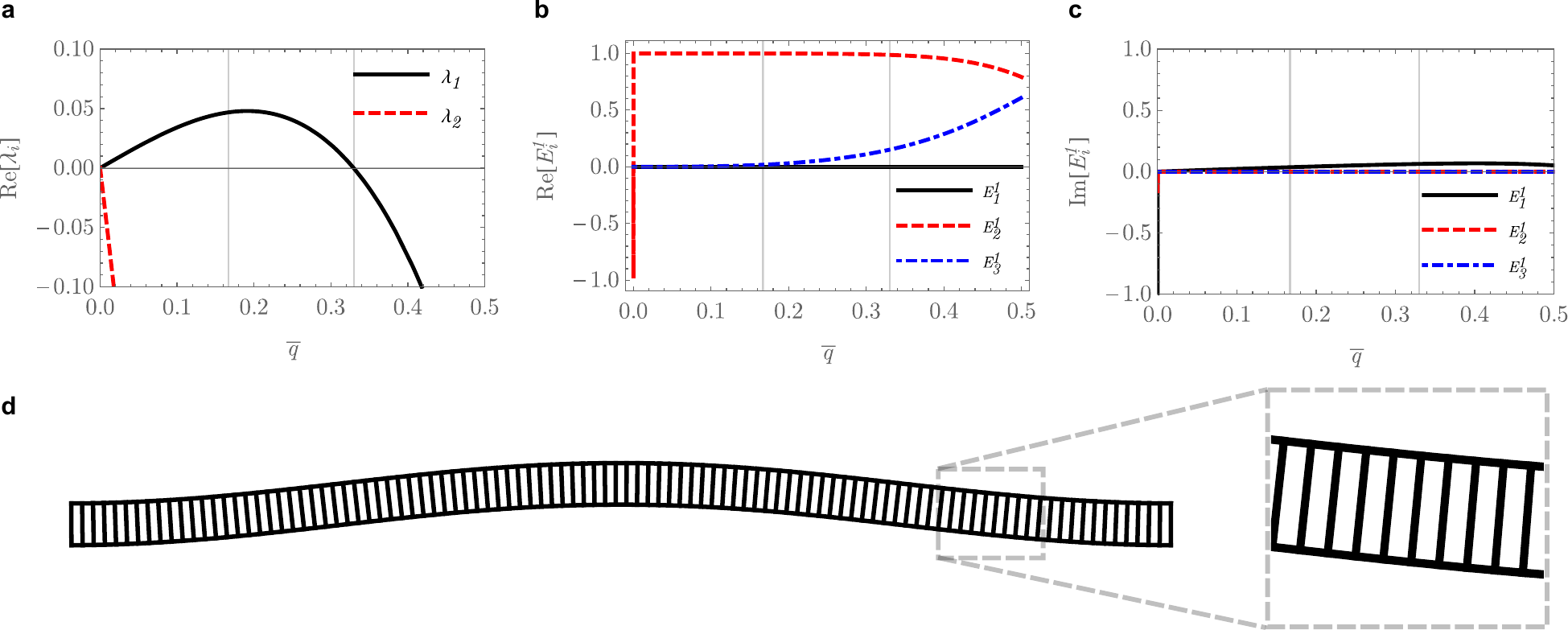}
%/home/rgm/Dropbox/Membranes/Epithelia/Elastomer/Figures/buck_lin_stab_sign.pdf
\caption
{
	(Color online).  Constrained-buckling: $\delta=\sqrt{2}$, 
	$\overline{\beta}=15$, $\overline{\kappa}=1/2$.  The 
	$\mathrm{Re}\left[\lambda_1\right]$ instability bounded [panel {\bf a})], 
	with $\mathrm{Re}\left[\lambda_1\right]\to 0$ as $\overline{q}\to 0$.  For 
	convenience, we focus on characteristic behaviour at $\overline{q}=1/6$.  
	Cosinusoidal variations in $\rho_\mathrm{b}(x,t)$ [panel {\bf b}) blue 
	dot-dashed line] are negligible, and hence this is a passive instability, 
	rather than an actively-driven one.  Since $\delta > 1$, the steady-state 
	corresponds to lateral faces that are already stretched.  As a result, 
	cosinusoidal variations in $h(x,t)$ [panel {\bf b}) red dashed line] 
	combined with negligible (sinusoidal) variations in $\omega(x,t)$ [panel 
	{\bf c}) black solid line] act to reduce lateral lengths at the expense of 
	apical expansion, due to the conservation of cell volume.
}
\label{fig:buck}
\end{figure*}

\end{widetext}

\end{document}